\documentclass[final]{siamart190516}
\usepackage[utf8]{inputenc}
\usepackage{graphicx}
\usepackage[nohead,margin=1.0in]{geometry}
\usepackage[tableposition=top]{caption}

\usepackage{amsmath}
\usepackage{commath}
\usepackage{amsfonts}
\usepackage{multirow}
\usepackage{makecell}
\usepackage{algorithmic}
\usepackage{algorithm}
\usepackage{tikz}
\usepackage{mathdots}
\usepackage{yhmath}
\usepackage{cancel}
\usepackage{array}
\usepackage{multirow}
\usepackage{amssymb}
\usepackage{gensymb}
\usepackage{tabularx}
\usepackage{booktabs}
\usetikzlibrary{fadings}
\usetikzlibrary{patterns}
\usetikzlibrary{shadows.blur}
\usetikzlibrary{shapes}

\usepackage{booktabs}
\usepackage{graphicx}
\usepackage{xcolor}
\usepackage{pgfplots}
\usepackage{siunitx}
\pgfplotsset{compat=1.8}
\usetikzlibrary{positioning,calc,arrows,arrows.meta,shapes,shapes.multipart,pgfplots.statistics}

\usepgfplotslibrary{statistics}
\pgfplotsset{compat=1.6}
\pgfplotsset{
	matrixaxis/.style={
    	scale only axis,
        view={0}{90},
        enlargelimits=false,
        axis on top,
        axis equal image,
		colormap/blackwhite,
        point meta max=0.5,
        point meta min=0.0,
        colorbar style={
        	yticklabel style={
                /pgf/number format/.cd,
                fixed,
                precision=2,
                fixed zerofill,
            },
            ylabel={NRMSE},
            ylabel style={
            	yshift=5mm,
                rotate=180,
            },
        },
    },
}

\newcommand{\N}{\mathbb{N}}

\newcommand{\R}{\mathbb{R}}

\DeclareMathOperator{\diva}{div}

\newtheorem{remark}{Remark}

\renewcommand{\d}{\, \text{d}}

\title{A simulation framework for particle magnetization dynamics of large ensembles of single domain particles: Numerical treatment of Brown/N\'{e}el dynamics and parameter identification problems in magnetic particle imaging\thanks{Submitted to the editors \today.
\funding{H. Albers and T. Kluth acknowledge funding by the German Research Foundation (DFG, Deutsche Forschungsgemeinschaft) - project 426078691. }}
}

\author{Hannes Albers\thanks{Center for Industrial Mathematics, University of Bremen, Bremen, Germany (\email{halbers@math.uni-bremen.de}, \email{tkluth@math.uni-bremen.de}).}
\and Tobias Kluth\footnotemark[2]
\and Tobias Knopp\thanks{Section for Biomedical Imaging, University Medical Center Hamburg-Eppendorf, Hamburg, Germany, and Institute for Biomedical Imaging, Hamburg University of Technology, Hamburg, Germany}
}

\begin{document}

\maketitle
\begin{abstract}
Magnetic nanoparticles and their magnetization dynamics play an important role in many applications. 
We focus on magnetization dynamics in large ensembles of single domain nanoparticles being characterized by either Brownian or N\'{e}el rotation mechanisms. 
Simulations of the respective behavior are obtained by solving advection-diffusion equations on the sphere, for which a unified computational framework is developed and investigated.
This builds the basis for solving two parameter identification problems, which are formulated in the context of the chosen application, magnetic particle imaging.
The functionality of the computational framework is illustrated by numerical results in the parameter identification problems either compared quantitatively or qualitatively to measured data. 

\end{abstract}



\section{Introduction}
%
%

Magnetic nanoparticles (MNPs) play an important role in biomedical applications and have been significantly researched during the past decades. Applications can be found in medical diagnosis and treatment. For instance magnetic particles are used for contrast enhancement in magnetic resonance imaging (MRI) \cite{bonnemain1998superparamagnetic} and for contrast generation in magnetic particle imaging (MPI) \cite{knopp2017magnetic}. In medical diagnosis, the particles generate a tissue-independent contrast and thus have many potential medical applications ranging from vascular imaging \cite{kaul2015combined,vaalma2017magnetic} to molecular imaging \cite{Zheng2015,
yu2017magnetic, arami2017tomographic}. Beside diagnosis, the particles can also be used for treatment, e.g., when using them as carriers for targeted drug delivery exploiting magnetic forces \cite{griese2020simultaneous} or when heating the particles in hyperthermia applications such as cancer treatment \cite{murase2015usefulness, sajjamark2020spatial, weber2020hyper}.

While for some applications the physical behavior of the particles needs to be known only qualitatively, there are several applications where a precise model for the particle physics is crucial. 
The magnetization dynamics for MNPs are originated in the field of micromagnetics \cite{brown1962} while individual MNPs are often modeled on  a larger scale in terms of their resulting magnetic moment \cite{Kluth2018a}. And for certain applications (e.g., MNP tracer) the behavior of large numbers of MNPs in an ensemble is of interest.
In this context, simplified models build on the theory of paramagnetism such that the mean magnetic moment of many MNPs is modeled explicitly based on the Langevin function \cite{rahmer2009signal,Knopp2010d,Goodwill2010,Kluth2017}. 
Nevertheless, this simple model relies on the assumption that a static magnetic field is applied and that the equilibrium state of the particles is reached at least in good approximation. 
However, these assumptions are in general not fulfilled for fast changing magnetic fields. 
The dynamic behavior of the MNPs' magnetic moments is affected by Brownian and N\'{e}el mechanisms (see also \cite{Coffey1992,shliomis1994theory}). The former describes the magnetic moment rotation due to rotation of the whole particle while the later describes the internal rotation of magnetic moment.

As an explicit example we focus on the imaging modality MPI which strongly relies on the change of the nanoparticles' magnetization in large ensembles of particles. 
In MPI the previously mentioned rotation dynamics were initially suggested in \cite{Weizenecker2010particle} in terms of stochastic ordinary differential equations. However, numerically solving these equations in order to get a good approximation of the mean requires significant computational power \cite{shasha2020}.
Methods for solving the Fokker-Planck equation for special cases like one-dimensional or spherically polarized magnetic fields were presented in \cite{Yoshida2012, Usadel2013, Fock2018}; see also \cite{weizenecker2012micro,Yoshida2012c,yoshida2013characterization,rogge2013simulation,reeves2014approaches,Kluth2018a} for further reading. 
The N\'{e}el rotation mechanism is further influenced by the particle anisotropy, which, for example, might be modeled by the orientation of the particle's easy axis in case of an uni-axial anisotropy. This can significantly affect the magnetization behavior of ensembles of MNPs \cite{Shah2015,Yoshida2017} and was recently exploited to formulate an approximate model for the MNP behavior for various offset fields \cite{KluthSzwargulskiKnopp2019}.
This specific magnetization behavior has motivated an increasing number of studies focusing on MPI-specific excitation patterns in Brownian and N\'{e}el rotations \cite{weizenecker2012micro,Martens2013,Deissler2014,Enpuku2014,Graeser2015c} and also the coupled Brown-N\'{e}el case \cite{Weizenecker2010particle,usov2012dynamics,Graeser2015c,Weizenecker2018}.
A first step to quantify the dominating mechanism with respect to the model parameters was made in \cite{reeves2015combined} motivating the consideration of either Brownian or N\'{e}el rotation in the context of parameter identification.

A more precise relationship between particle physics and the desired application is presented in the following description of a general MPI experiment: MNPs (contained in a tracer material) are located in a MPI scanner and a dynamic applied magnetic field causes a change of their magnetization inducing a voltage $\tilde{v}^\mathrm{P}:[0,T] \rightarrow \R$ in so-called receive coil units characterized by a sensitivity profile $\mathbf{p}^R: \R^3 \rightarrow \R^3$, i.e.,
\begin{align}
 \tilde{v}^\mathrm{P} (t) &=- \mu_0 \int_\Omega \mathbf{p}^R(x) \cdot \frac{\partial}{\partial t} \mathbf{M} (x,t)  \d x = \int_\Omega c(x) \underset{=s(x,t)}{\underbrace{(-\mu_0\mathbf{p}^R(x) \cdot \frac{\partial}{\partial t} \bar{\mathbf{m}} (x,t))}}  \d x,  \label{eq:MPI_static_concentration}
\end{align}
where $c: \Omega \rightarrow \R^+_0$ is the concentration of the MNPs, $\mathbf{\bar{m}}: \R^3 \times [0,T] \rightarrow \R^3$ is the resulting mean magnetic moment of the particles and $\Omega\subset\R^3$ is the imaging volume. The approximation of the magnetization $\mathbf{M}$ in terms of the concentration and the mean magnetic moment can be derived by doing a transition from microscopic to macroscopic scale (see \cite{Kluth2018a} for further details).
The mean magnetic moment $\mathbf{\bar{m}}(x,t)$ depends on the applied magnetic field $\mathbf{H}: \R^3 \times [0,T] \rightarrow \R^3$, which is usually a $\mathfrak{T}$-periodic function with period $\mathfrak{T}$ along the time dimension (often the case $T=\mathfrak{T}$ or $T=k\mathfrak{T}$, $k\in \N$, is considered). In MPI a static selection field $\mathbf{H}_\mathrm{S}: \R^3 \rightarrow \R^3$ is combined with a dynamic and approximately homogeneous drive field $\mathbf{H}_\mathrm{D}: [0,T]\rightarrow \R^3$, i.e., $\mathbf{H}(x,t)=\mathbf{H}_\mathrm{S}(x) + \mathbf{H}_\mathrm{D}(t)$. 
From a local perspective, the selection field can be seen as a set of offset fields $h_\mathrm{S} \in \R^3$ varying within the field-of-view and thus encoding the space-dependent information of the tracer concentration. 
Note that the computational framework outlined in this work is thus developed for general applied fields $\mathbf{H}(t)=\mathbf{h}_\mathrm{D}(t) + h_\mathrm{S}$  for fixed offset field $h_\mathrm{S}\in \R^3$ and dynamic fields $\mathbf{h}_\mathrm{D}: [0,T] \rightarrow \R^3$ such that a spatial dependence (with respect to the variable $x$) is omitted in the computational simulation framework.

In MPI not only the particle magnetization but also the applied field $\mathbf{H}$ induces a voltage $\tilde{v}^\mathrm{E}(t)$, which is known as direct feedthrough.
Since this value is several orders of magnitude larger than the particle signal, it must be removed prior to digitization. This is done by an analog filter in the analog signal  chain, which is formally represented by a convolution with a $\mathfrak{T}$-periodic filter kernel $a:\R \rightarrow \R$ yielding the signal $v=a\ast(\tilde{v}^\mathrm{P}+\tilde{v}^\mathrm{E})$.
The resulting integral kernel (also called \emph{system function}; \emph{system matrix} in the discretized setup) is $\tilde{s}(x,t)=(s(x,\cdot)*a)(t)$.
As already proposed in the initial MPI publication \cite{Gleich2005}, the system function is typically determined in a time-consuming measurement-based calibration process which additionally suffers from limited generalizability with respect to device as well as tracer parameters.
An alternative approach relies on proper modeling the mean magnetic moment $\mathbf{\bar{m}}$ of large ensembles of nanoparticles and
the identification of certain model parameters.
This is still one of the unsolved open problems in MPI research, which is partially addressed in the present work.

In the remainder of the work we focus on computational aspects when modeling the behavior of large ensembles of nanoparticles in a dynamic applied magnetic field and parameter identification problems for the purpose of model-based calibration. 
The work is structured as follows.
In Section \ref{sec:magnetization_models} we briefly introduce the relevant magnetization dynamic model for large ensembles of nanoparticles exploiting a Fokker-Planck equation to include stochastic behavior of individual nanoparticles.
Algorithmic solutions for this type of equations and numerical results on computational performance are presented in Sections \ref{sec:computational_framework} and \ref{sec:comp_eval} and can be found in a toolbox at \url{https://github.com/MagneticParticleImaging/MNPDynamics}.
Two parameter identification problems in the context of model-based system calibration in MPI are discussed in Section \ref{sec:parameter_identification}. 
The first example is a quantitative polydisperse model fit to magnetic particle spectrometer (MPS) measurements of immobilized and oriented tracer samples. 
The second example is a simulation-based determination of a convolution kernel as used in the so-called $x$-space method \cite{goodwill2010x, Goodwill2011} which qualitatively mimics the observed behaviour for pulsed as well as sinusoidal excitation measured data in \cite{tay2019pulsed}. 
We conclude with a discussion in Section \ref{sec:discussion}.

\begin{remark}
Please note that in the preceding part and in the remainder of this work vector-valued functions are in bold font, e.g., $\mathbf{\bar{m}}$ and vector-valued function \textit{arguments} are non-bold, e.g. the space variable $x$ and the variable $m$ in the following.
\end{remark}

\section{Magnetization in large particle ensembles}
\label{sec:magnetization_models}


We consider large ensembles of single domain nanoparticles with uni-axial anisotropy whose magnetization behavior is mainly determined by two dynamic mechanisms: the Brownian
rotation describes the mechanical alignment of the entire particle with a change of the magnetic field whereas N\'eel rotation describes the alignment of the particle’s inner magnetic moment. 
Both dynamic mechanisms can be described by Langevin equations for individual particles taking into account thermal noise (see \cite{Kluth2018a} for a survey). 
This starting point allows for two different approaches to obtain an estimate for the mean magnetic moment of a large ensemble. One can either consider the
problem for individual particles and solve the Langevin equation for a sufficiently large number of particles to obtain a reasonable estimate for
the mean. Alternatively, one can take a comprehensive view and solve the Fokker–Planck equation for a
probability distribution representing an entire ensemble of nanoparticles in terms of a parabolic partial
differential equation.
Following the latter approach we can comprise these two dynamic mechanisms in the following advection-diffusion model, which is at the core of the computational simulation framework outlined and exploited in this work. For details on the rotation models in terms of the Langevin equation and the derivation of the corresponding Fokker Planck equation we refer to the survey \cite{Kluth2018a}.
In the remainder of this work let $S:=S^2\subset \R^3$ denote the surface of the sphere.

\begin{definition}[Advection-diffusion rotation model]
 \label{def:MPI-model-advection-diffusion}
The mean magnetic moment vector
\begin{equation}
 \mathbf{\bar{m}}(t) = m_0 \int_{S} m f(m,t) \d m
\end{equation}
for $t\in I:=[0,T]$ and applied field $\mathbf{H}(t)=\mathbf{h}_\mathrm{D}(t) + h_\mathrm{S}$, $\mathbf{h}_\mathrm{D}:I \rightarrow \R^3$, $h_\mathrm{S} \in \R^3$, is given in terms of the probability density function $f: S \times I \rightarrow \R^+\cup \{0\}$, which is the solution to
\begin{equation}
\label{eq:FP-general}
\left\{
\begin{aligned}
&\begin{aligned} \frac{\partial}{\partial t} f = \mathrm{div}_{S}(\frac{1}{2\tau} \nabla_{S} f ) - \mathrm{div}_{S}(\mathbf{\tilde{b}} f) \end{aligned} & & \text{in } S\times I \\
&f(\cdot,0)=f_0  & &  \text{in } S
\end{aligned}
 \right.
\end{equation}
where $\tau >0$ is the relaxation time constant, $f_0: S \rightarrow \R^+ \cup \{0\}$ with $\int_{S} f_0 \ dm=1$ is the initial distribution function, and the (velocity) field $\mathbf{\tilde{b}}:S \times \R^3 \times S \rightarrow \R^3$ given by
\begin{align}
 &\mathbf{\tilde{b}}(m,\mathbf{H},n) = p_1 \mathbf{H} \times m + p_2 (m\times \mathbf{H}) \times m \notag \\ & \quad+ p_3 (n\cdot m) n \times m + p_4 (n\cdot m) (m\times n) \times m \label{eq:advection}
\end{align}
where $p_i\geq 0$, $i=1,\hdots,4$, are physical constants and $n\in S$ is the easy axis of the particles.
\end{definition}

 A pure N\'{e}el rotation including anisotropy is given by $p_1=\tilde{\gamma}\mu_0$, $p_2=\tilde{\gamma}\alpha \mu_0$, $p_3=2\tilde{\gamma}\frac{K_\mathrm{anis}}{M_\mathrm{S}}$, $p_4=2\tilde{\gamma}\alpha\frac{K_\mathrm{anis}}{M_\mathrm{S}}$, and $\tau=\frac{V_\mathrm{C} M_\mathrm{S}}{2 k_\mathrm{B} T_\mathrm{B} \tilde{\gamma} \alpha}$ ($\tilde{\gamma}=\frac{\gamma}{1+ \alpha^2}$).
 The Brownian case is covered by the parameter set $p_2=\mu_0\frac{V_\mathrm{C} M_\mathrm{S}}{6 \eta V_\mathrm{H}}$, $p_1=p_3=p_4=0$, and $\tau=\frac{3 V_\mathrm{H}\eta}{k_\mathrm{B} T_\mathrm{B}}$. These quantities denote the volume of the nanoparticle's magnetizable core $V_\mathrm{C}$, the nanoparticle's hydrodynamic volume $V_\mathrm{H}$, viscosity $\eta$, anisotropy constant $K_\mathrm{anis}$, saturation magnetization $M_\mathrm{S}$, gyromagnetic ration $\gamma$, damping constant $\alpha$, Boltzmann constant $k_\mathrm{B}$, temperature $T_\mathrm{B}$, and magnetic permeability in vacuum $\mu_0$.  

\section{Algorithms}
\label{sec:computational_framework}
We present two approaches to discretize the Fokker-Planck equation in \eqref{eq:FP-general} at hand, which are both comprised in the computational simulation framework/toolbox. Both rely on the \emph{method of lines}, i.e., the discretization is carried out with respect to the spatial variable (here $m$) only while the time variable is left continuous. This results in a system of ODEs, the number of equations depending on the fineness of the spatial approximation. For more information regarding this approach, see e.g. \cite{schiesser2009}.

Since the approaches outlined in this section are for the most part not specific to the Fokker-Planck equation, we shall consider a general advection-diffusion equation defined on the sphere $S$ for some time interval $I\subset \R$, advection term $\mathbf{b}: S \times I \rightarrow \R^3$ and diffusion constant $c>0$, denoting the unknown solution as $u: S \times I \rightarrow \R$ adopting the typical notation in the computational PDE literature within this section:
\begin{equation}\label{eq:general_adv_diff}
\frac{\partial u}{\partial t} = \diva \left( -\mathbf{b}u + c \nabla u\right) \quad \text{ in } S\times I.
\end{equation}
\subsection{Approach A: Spherical Harmonics} The first approach is a so-called \emph{Galerkin method} utilizing the \emph{variational formulation} of the PDE. For a comprehensive overview, see e.g. \cite{quarteroni}.

In a nutshell, the variational formulation of a given PDE can be obtained by multiplying it with an arbitrary test function and integrating the resulting equation. Then, integration by parts is used to reformulate the second-order PDE into a variational equation involving only first-order derivatives. What results is a bilinear (or in the complex case sesquilinear) form $A:V \times V \to \mathbb{K}$ for $\mathbb{K}\in \{\mathbb{R}, \mathbb{C}\}$, defined on a suitable function space $V$, in our case the Sobolev space $H^1(S)$.

For the Galerkin method, finite-dimensional subspaces $V_h \subset V$ are considered with $\inf_{v_h \in V_h} ||v_h-v|| \to 0$ as $h \to 0$ for each $v \in V$ and the finite-dimensional restriction of the bilinear form $A$ to $V_h$ is used to obtain approximations. For suitable assumptions on $A$, the solutions $u_h$ to these restricted problems can be shown to tend to the solution $u$ of the full variational problem as $h \to 0$ \cite{quarteroni}.

Since the solution as well as the test functions are defined on the sphere, a canonical choice for the subspaces $V_h$ are the spherical harmonic (SH) functions: $V_{1/N}:= V^{SH}_N := \operatorname{span}\{Y^l_m: l \leq N, m = -l,\dots,l\}$.

The (non-normalized) spherical harmonics are functions on $S$ and can be represented as
\begin{align*}
Y_m^l(\theta, \phi) = P_{lm}(\cos \theta) e^{im\phi}
\end{align*}
where $P_{lm}$ are the associated Legendre polynomials. These functions are smooth, i.e. $V_N^{SH} \subset V = H^1(S)$, and form an orthogonal basis of $L^2(S)$ \cite{nolting2017theoretical}.\\
In order to obtain the discrete linear operator $\hat{A}$ for this choice of approximation spaces, we need to compute the sesquilinear forms evaluated on the basis functions: $A(\varphi_j, \varphi_i) = A(Y_{m_j}^{l_j}, Y_{m_i}^{l_i})$ for $i,j = 1, \dots,(N+1)^2 = \dim V_N^{SH} $. For simplicity, we will sometimes write $Y_{m_k}^{l_k}$ as $Y_k$ for short. This yields:
\begin{align}\label{shintegrals}
    A(Y_j, Y_i) &= c\int_S \nabla Y_j^* \cdot\nabla  Y_i \, dx + \int_S (\mathbf{b} \cdot \nabla Y_i) Y_j^* \, dx \nonumber \\
    & = -c \int \Delta Y_j^* Y_i  \, dx + \int (\mathbf{b} \cdot \nabla Y_i) Y_j^* \, dx \nonumber \\
    &= c \,l_j(l_j+1) \int Y_j^* Y_i \, dx + \int (\mathbf{b} \cdot \nabla Y_i) Y_j^* \, dx \nonumber \\
    &= c \, C_j l_j(l_j+1) \delta_{ij} + \int_S (\mathbf{b} \cdot \nabla Y_i) Y_j^* \, dx.
\end{align}
Since the spherical harmonics are not normalized in our case, we have that $\int Y_i Y_j^*\, dx = C_j \delta_{ij}$ for some $C_j > 0$.

Calculating the first term in \eqref{shintegrals} is straightforward and requires no computational effort. The second term, however, is significantly harder to evaluate. For general $\mathbf{b}$, we need to numerically calculate an integral over the sphere, where the integrand is nonzero for the most part. Since we still have time as a parameter, these values would have to be recalculated for each timestep, since $\mathbf{b}$ is time-dependent. Also, the resulting matrix $\hat{A}:=(A(Y_j,Y_i))_{i,j}$ is not sparse, making the integration of the ODE harder. Since $A$ is a non-symmetric sesquilinear form, we would need to calculate $(N+1)^4$ of these integrals per timestep, making this approach prohibitively expensive.

However, for the special form of the advection term $\mathbf{b}$ that we have for the Fokker-Planck problem, i.e., $\mathbf{b}=\mathbf{\tilde{b}}(m,\mathbf{H},n)$ for given $\mathbf{H}$ and $n$, the calculation of the entries of $\hat{A}$ can be simplified significantly by rewriting $\mathbf{b} \cdot \nabla Y_k$ as an operator acting on $Y_k$, such that the result can be expressed exclusively in terms of other SH functions $Y_i$. This process is detailed in Appendix \ref{sec:appendix_SH} and yields an ODE system of the form 
\begin{align}\label{odegeneralform}
    \frac{\partial C^q_r}{\partial t} = \sum_{q'=q-2}^{q+2}\sum_{r'=r-2}^{r+2}\gamma^{q,q'}_{r,r'}(t) C^{q'}_{r'}(t), \qquad q=-r,\dots,r,\quad r=0,\dots,\infty,
\end{align}
where $C^q_r$ are the SH coefficients of the probability density function, i.e.
\begin{align*}
u(m,t) = \sum_{q,r} C^q_r(t) Y^q_r(m)
\end{align*}
and $\gamma^{q,q'}_{r,r'}$ are coefficients depending on the applied magnetic field and the physical constants $p_i,  i=1,\dots,4$. The coefficients are listed in Appendix \ref{sec:appendix_SH} and are applicable for time-dependent magnetic field $\mathbf{H}$ as well as time-dependent easy axis $n$. The derivation follows that in \cite{Weizenecker2018}, but has been extended to allow for time-dependent easy axes. For a fixed easy axis, the system can be simplified by rotating the coordinate system such that the easy axis aligns with the $z$-axis, eliminating many of the terms.

\subsubsection{Algorithm A: Constructing the SH matrix}
Since we only computed the discretization of the right-hand side of the PDE, we are left with a system of ODEs with $(N+1)^2$ equations of the form
\begin{align}\label{eq:general_disc}
    \Dot{\xi}(t) = M(t)\xi(t), \quad \xi(t) \in \mathbb{R}^{(N+1)^2},\,t \in I.
\end{align}
Here, $\xi$ is the vector of SH coefficient, i.e. $\xi(t) = (C^0_0(t),\dots,C^{q_\mathrm{max}}_{r_\mathrm{max}}(t))$, and the matrix $M$ comprises the time-dependent coefficients as introduced in equation \eqref{odegeneralform}. For solving this system, one of the many ODE integrators for stiff systems can be used. We use the \texttt{ode15s}-routine from \textsc{Matlab}.\\
Such a method only requires an initial value $\xi_0$ and a function that outputs $M(t)$ at each time $t$ that the integrator requires (these integrators typically work with variable timesteps where every step length is computed based on the current solution).\\
As the initial value, we commonly choose a uniform probability distribution (all directions of the magnetic moment vector are equally likely), which corresponds to $C^0_0=1/4\pi$ and $C^q_r = 0$ for $(r,q)\neq (0,0)$. Thus, $\xi_0 = (1/4\pi)e_1$.\\
The matrix $M(t)$ has to be reevaluated in each time step. However, since only $\mathbf{H}$ and (possibly) $n$ are time dependent, computation can be optimized to minimize complexity. Algorithm \ref{algomatrix} describes how $M(t)$ is assembled, given a function for the coefficients $\gamma^{q,q'}_{r,r'}$ depending on $\mathbf{H}(t)$ and $n(t)$.
\begin{algorithm}
\caption{Computing the matrix $M(t^*)$ for a fixed time $t^*$.}\label{algomatrix}
\begin{algorithmic}
\STATE Given coefficients $\{\gamma^{q,q'}_{r,r'}(t^*)\}$ for fixed $t^*$ and zero matrix $M \in \mathbb{R}^{(N+1)^2\times (N+1)^2}$
    \FOR{$i=1,\dots,(N+1)^2$}
        \STATE convert index $i$ to quantum numbers $(l_i,m_i)$
        \FOR{$r=l_i-2,\dots,l_i+2$}
        \FOR{$q=m_i-2,\dots,m_i+2$}
        \IF{$(r,q)$ are valid quantum numbers}
        \STATE convert quantum numbers $(r,q)$ to index $j$
        \STATE \textbf{set} $M(i,j) = \gamma^{m_i,q}_{l_i,r}(t^*)$
        \ENDIF
        \ENDFOR
        \ENDFOR
        \ENDFOR
        \RETURN $M$
\end{algorithmic}
\end{algorithm}
\subsection{Approach B: The finite volume method}\label{sectionFV}
The finite volume (FV) method is a different method of discretization, where a triangular mesh in space (more precisely: on the sphere) is considered. Assume we have a triangulation of the sphere that decomposes it into spherical triangles $T_i$, $i= 1,\dots,N$. We again examine the general advection-diffusion equation \eqref{eq:general_adv_diff}.

\begin{figure}[ht]
\centering
\tikzset{every picture/.style={line width=0.75pt}} 
\begin{tikzpicture}[x=0.75pt,y=0.75pt,yscale=-.5,xscale=.5]

\draw   (227.98,113.76) -- (472.06,584.98) -- (104.76,450.64) -- cycle ;
\draw   (227.98,113.76) -- (472.06,584.98) -- (541.16,16.43) -- cycle ;
\draw  [fill={rgb, 255:red, 0; green, 0; blue, 0 }  ,fill opacity=1 ] (319.13,359.1) .. controls (319.13,356.84) and (320.97,355) .. (323.23,355) .. controls (325.5,355) and (327.33,356.84) .. (327.33,359.1) .. controls (327.33,361.36) and (325.5,363.2) .. (323.23,363.2) .. controls (320.97,363.2) and (319.13,361.36) .. (319.13,359.1) -- cycle ;
\draw  [fill={rgb, 255:red, 0; green, 0; blue, 0 }  ,fill opacity=1 ] (457.8,277.43) .. controls (457.8,275.17) and (459.64,273.33) .. (461.9,273.33) .. controls (464.16,273.33) and (466,275.17) .. (466,277.43) .. controls (466,279.7) and (464.16,281.53) .. (461.9,281.53) .. controls (459.64,281.53) and (457.8,279.7) .. (457.8,277.43) -- cycle ;
\draw  [fill={rgb, 255:red, 0; green, 0; blue, 0 }  ,fill opacity=1 ] (343.13,343.1) .. controls (343.13,340.84) and (344.97,339) .. (347.23,339) .. controls (349.5,339) and (351.33,340.84) .. (351.33,343.1) .. controls (351.33,345.36) and (349.5,347.2) .. (347.23,347.2) .. controls (344.97,347.2) and (343.13,345.36) .. (343.13,343.1) -- cycle ;
\draw    (325.43,358.11) -- (346.75,344.29) ;
\draw [shift={(348.42,343.2)}, rotate = 507.03] [color={rgb, 255:red, 0; green, 0; blue, 0 }  ][line width=0.75]    (10.93,-3.29) .. controls (6.95,-1.4) and (3.31,-0.3) .. (0,0) .. controls (3.31,0.3) and (6.95,1.4) .. (10.93,3.29)   ;
\draw [shift={(323.76,359.2)}, rotate = 327.03] [color={rgb, 255:red, 0; green, 0; blue, 0 }  ][line width=0.75]    (10.93,-3.29) .. controls (6.95,-1.4) and (3.31,-0.3) .. (0,0) .. controls (3.31,0.3) and (6.95,1.4) .. (10.93,3.29)   ;
\draw    (349.49,342.2) -- (459.69,278.53) ;
\draw [shift={(461.42,277.53)}, rotate = 509.98] [color={rgb, 255:red, 0; green, 0; blue, 0 }  ][line width=0.75]    (10.93,-3.29) .. controls (6.95,-1.4) and (3.31,-0.3) .. (0,0) .. controls (3.31,0.3) and (6.95,1.4) .. (10.93,3.29)   ;
\draw [shift={(347.76,343.2)}, rotate = 329.98] [color={rgb, 255:red, 0; green, 0; blue, 0 }  ][line width=0.75]    (10.93,-3.29) .. controls (6.95,-1.4) and (3.31,-0.3) .. (0,0) .. controls (3.31,0.3) and (6.95,1.4) .. (10.93,3.29)   ;
\draw  [dash pattern={on 0.84pt off 2.51pt}]  (352,428.67) -- (347.09,343.2) ;
\draw    (261,176.33) -- (330.97,141.43) ;
\draw [shift={(332.76,140.53)}, rotate = 513.48] [color={rgb, 255:red, 0; green, 0; blue, 0 }  ][line width=0.75]    (10.93,-3.29) .. controls (6.95,-1.4) and (3.31,-0.3) .. (0,0) .. controls (3.31,0.3) and (6.95,1.4) .. (10.93,3.29)   ;
\draw   (268.13,172.65) -- (271.69,179.66) -- (264.67,183.22) -- (261.11,176.21) -- cycle ;

\draw (208,242.07) node [anchor=north west][inner sep=0.75pt]  [font=\LARGE]  {$T_{i}$};
\draw (245.33,516.07) node [anchor=north west][inner sep=0.75pt]    {$E_{i_1}$};
\draw (300,230) node [anchor=north west][inner sep=0.75pt]    {$E_{i_2}$};
\draw (110,282.73) node [anchor=north west][inner sep=0.75pt]    {$E_{i_3}$};
\draw (435.33,109.4) node [anchor=north west][inner sep=0.75pt]  [font=\LARGE]  {$T_{i_2}$};
\draw (300,310) node [anchor=north west][inner sep=0.75pt]  [font=\footnotesize]  {$h_{i_2}$};
\draw (378.67,270) node [anchor=north west][inner sep=0.75pt]  [font=\footnotesize]  {$\overline{h_{i_2}}$};
\draw (335,433.73) node [anchor=north west][inner sep=0.75pt]  [font=\footnotesize]  {$E_{i_{2_{\mathrm{mid}}}}$};
\draw (264,130) node [anchor=north west][inner sep=0.75pt]  [font=\footnotesize] [align=left] {$\displaystyle \mathbf{e}_{i_2}$};
\end{tikzpicture}
\caption{Triangle $T_i$ and one of its adjacent triangles with quantities required for the Finite Volume method.}\label{FVillustration}
\end{figure}
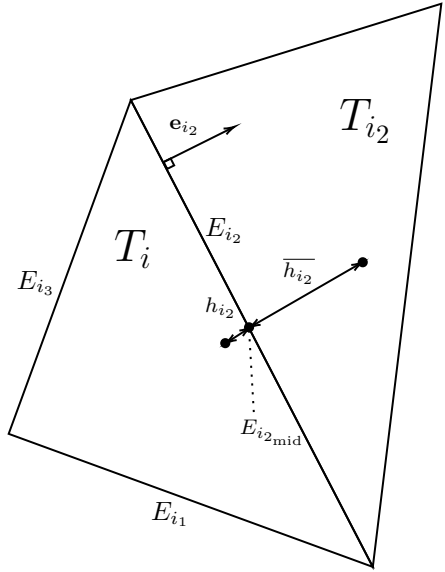

We consider a discretization $u(x) \approx u_i$ for $x \in T_i$, where $u_i$ is the mean value of $u$ on the $i$-th triangle, i.e. $u_i = \frac{1}{\abs{T_i}} \int_{T_i} u(x) \, dx$. Thus, we obtain the equations
\begin{align*}
\frac{\partial }{\partial t} u_i = \frac{1}{\abs{T_i}} \int_{T_i} \diva(-\mathbf{b} u) \, dx + \frac{1}{\abs{T_i}} \int_{T_i}c\Delta u \, dx, \quad i=1,\dots,N.
\end{align*}
By applying the divergence theorem and using the fact that a triangle's border is the union of its edges, i.e. $\partial T_i = E_{i_1} \cup E_{i_2} \cup E_{i_3}$, we get
\begin{align*}
\frac{\partial u_i}{\partial t} &= -\frac{1}{\abs{T_i}} \int_{\partial T_i} \mathbf{b}u \cdot \mathbf{e}_i \, d\sigma + \frac{1}{\abs{T_i}}c\int_{\partial T_i} \nabla u \cdot \mathbf{e}_i \, d\sigma\\
&= -\frac{1}{\abs{T_i}} \sum_{i=1}^3 \left[ \int_{E_{i_j}} \mathbf{b}u \cdot \mathbf{e}_{i_j} \, d \sigma -c \int_{E_{i_j}} \nabla u \cdot \mathbf{e}_{i_j} \, d \sigma \right],
\end{align*}
where $\mathbf{e}_{i_j}$ is the outer surface normal of the $j$'th edge of the $i$'th triangle.\\
Let us first consider the first integral that contains the advection term $\mathbf{b}$. We approximate it using the midpoint-rule:
\begin{align*}
\int_{E_{i_j}} \mathbf{b}u \cdot \mathbf{e}_{i_j} \, d \sigma \approx \mathbf{b}(E_{i_{j_{\mathrm{mid}}}}) u(E_{i_{j_{\mathrm{mid}}}}) \cdot \mathbf{e}_{i_j} \abs{E_{i_j}},
\end{align*}
where now $E_{i_{j_{\mathrm{mid}}}}$ is the midpoint of the $j$'th edge of the $i$'th triangle.\\
Since the discretized function is in general not continuous at the edges (since it's defined to be piecewise constant on each triangle), we approximate $u(E_{i_{j_{\mathrm{mid}}}})$ by a weighted average of the discretized function's value on the two triangles, weighted by the distance of the edge's midpoint to the two triangle circumcenters. Using the circumcenter instead of the centroid of the triangle simplifies calculations, but imposes a restriction on the triangulation, since the circumcenter of each triangle has to be in its interior.\\
Due to the properties of the finite volume method, this choice of weighting may lead to inaccurate results or reduced numerical stability if the advection term is of much greater magnitude than the diffusive term \cite{cfdbook}. An option to remedy this is to use so-called \emph{upwind discretization} instead, where in place of the average, only the node in upstream direction of the flow is taken into account. In our toolbox, we include the option to use a combination of upstream and central difference discretization, controlled by a parameter $\beta$, where $\beta = 0$ corresponds to only central differences and $\beta = 1$ corresponds to only upwind discretization. This follows the approach in \cite{zharovsky2012}.\\
For the central difference scheme, we denote the triangle that borders on triangle $i$'s $j$'th edge by $T_{i_j}$, defining $h_{i_j}$ to be the distance of $T_i$'s circumcenter to $E_{i_{j_{\mathrm{mid}}}}$, and $\overline{h_{i_j}}$ to be the distance of $T_{i_j}$'s circumcenter to the same point. For an illustration, see Figure \ref{FVillustration}.\\
This yields
\begin{align*}
\int_{E_{i_j}} \mathbf{b}u \cdot \mathbf{e}_{i_j} \, d \sigma &\approx \left[ \alpha_{i_j}u_i + (1-\alpha_{i_j})u_{i_j} \right] \mathbf{b}(E_{i_{j_{\mathrm{mid}}}})\cdot \mathbf{e}_{i_j}\abs{E_{i_j}}, \quad \text{with }\\
\alpha_{i_j} &= \frac{h_{i_j}}{h_{i_j}+\overline{h_{i_j}}}.
\end{align*}
With this, we can construct a matrix $A$ for the advection term as follows, where $\hat{u} = (u_1, \dots, u_N)$ is the discretized version of $u$:
\begin{align*}
\frac{1}{\abs{T_i}} \int_{T_i} \diva(-\mathbf{b} u) \, dx &\approx (A\hat{u})_i,\\
A_{ii} &= \frac{1}{\abs{T_i}}(\alpha_{i_1} d_{i_1}  + \alpha_{i_2} d_{i_2} + \alpha_{i_3} d_{i_3}),\\
A_{ii_j} &= \frac{1}{\abs{T_i}}[(1-\alpha_{i_j})d_{i_j}],\\
A_{ik} &= 0 \quad \text{ if triangle $k$ does not share an edge with triangle $i$},
\end{align*}
where $d_{i_j} = \mathbf{b}(E_{i_{j_{\mathrm{mid}}}})\cdot \mathbf{e}_{i_j}\abs{E_{i_j}}$.
Again, the index $i_j$ denotes the triangle index that borders on triangle $i$'s $j$'th edge.\\
The same thing can be done with the second integral, representing the diffusive term. In this case, the directional derivative in normal direction is replaced by a finite difference:
\begin{align*}
    c\int_{E_{i_j}} \nabla u \cdot \mathbf{e}_{i_j} \, d\sigma &\approx \nabla u(E_{i_{j_\mathrm{mid}}})\cdot \mathbf{e}_{i_j} \abs{E_{i_j}}\\
    & \approx \frac{u_{i_j}-u_i}{h_{i_j}+\overline{h_{i_j}}}\abs{E_{i_j}}.
\end{align*}
This yields a matrix $C$ representing the diffusive term:
\begin{align*}
    \frac{c}{\abs{T_i}}\int_{T_i} \Delta u \, dx & \approx (C\hat{u})_i,\\
    C_{ii} & = -\frac{c}{\abs{T_i}} \left( \frac{\abs{E_{i_1}}}{h_{i_1} + \overline{h_{i_1}}} + \frac{\abs{E_{i_2}}}{h_{i_2} + \overline{h_{i_2}}} + \frac{\abs{E_{i_3}}}{h_{i_3} + \overline{h_{i_3}}} \right), \\
    C_{ii_j} & = \frac{c}{\abs{T_i}} \frac{\abs{E_{i_j}}}{h_{i_j}+\overline{h_{i_j}}}, \\
    C_{ij} &= 0 \quad \text{ if triangle $j$ does not share an edge with triangle $i$.}
\end{align*}
Importantly, this matrix $C$ does not depend on time but depends only on the chosen triangulation. Thus, it can be computed in advance and be stored along with the other triangulation data.

\subsubsection{Algorithm B: Constructing the FV matrix}
For this approach, we assume that a discretization of the sphere into $N$ triangles is given, along with corresponding data such as the length of each edge, the outer normal on the edges, the position of the triangle circumcenters and many more. We further assume that the circumcenter of each triangle lies in its interior. Then, we again get an ODE system with $N$ equations in the form of \eqref{eq:general_disc}.
How this matrix $M$ can be obtained at each point in time has been described in Section \ref{sectionFV}, and is summarized in Algorithm \ref{algFV}.

\begin{algorithm}
\caption{Computing the matrix $M(t^*)$ for a fixed time $t^*$.}
\begin{algorithmic}\label{algFV}
\STATE given triangles $\{T_i\}$ and their areas $\{|T_i|\}$, edge normals $\{\mathbf{e}_{i_j}\}$, edge lengths $\{|E_i|\}$, edge centers $\{\mathbf{E}_{i_\mathrm{mid}}\}$ distance weights $\{\alpha_{i_j}\}$, advection function $\mathbf{b}(\cdot, t^*)$, upwind strength $\beta \in [0,1]$ and zero matrix $M\in \mathbb{R}^{N\times N}$
\FOR{$i = 1,\dots,N$ (triangle index)}
\FOR{$j=1,2,3$ (edge index)}
\STATE determine index $i_j$ of triangle $i$'s neighbor at edge $j$
\STATE calculate $d_{i_j} := \mathbf{b}(\mathbf{E}_{{i_j}_\mathrm{mid}})\cdot \mathbf{e}_{i_j}|E_{i_j}|$
\STATE set $\widehat{\alpha_{i_j}} = \beta\max(d_{i_j},0) + (1-\beta) \alpha_{i_j}d_{i_j}$
\STATE set $M(i,i_j) := \frac{1}{|T_i|}\left(1-\widehat{\alpha_{i_j}}\right)$
\ENDFOR
\STATE set $M(i,i) := \frac{1}{|T_i|}(\widehat{\alpha_{i_1}} + \widehat{\alpha_{i_2}} + \widehat{\alpha_{i_3}})$
\ENDFOR
\end{algorithmic}
\end{algorithm}

\section{Computational evaluation}\label{sec:comp_eval}
In this section, evaluations of the previously described algorithms will be presented. We show how computation times can be sped up by neglecting the precession term for many parameter settings, as well es considering the accuracy of different discretizations. Finally, we compare the computational results to the physical reality by examining two parameter identification results in the following Section \ref{sec:parameter_identification}.

\subsection{About the precession term}
We recall that for N\'{e}el relaxation, the advection term $\mathbf{b}$ has the form
\begin{equation*}
\mathbf{{b}}(m,t) = p_1 \mathbf{H}(t) \times m + p_2 (m\times \mathbf{H})(t) \times m + p_3 (n\cdot m) n \times m + p_4 (n\cdot m) (m\times n) \times m
\end{equation*}
We now want to consider the role of the terms pertaining to $p_1$ and $p_3$. In the Landau-Lifshitz-Gilbert equation that this expression is derived from, these terms lead to a precession of the magnetic moment around the magnetic field vector, which is why we call them \emph{precession terms}. During simulations for the mean magnetic moment, we have observed that neglecting these terms often leads to nearly identical results, but the computation times can be much faster.

\begin{figure}[h]
    \centering
    \begin{minipage}{0.49\linewidth}
    \includegraphics[width=\linewidth]{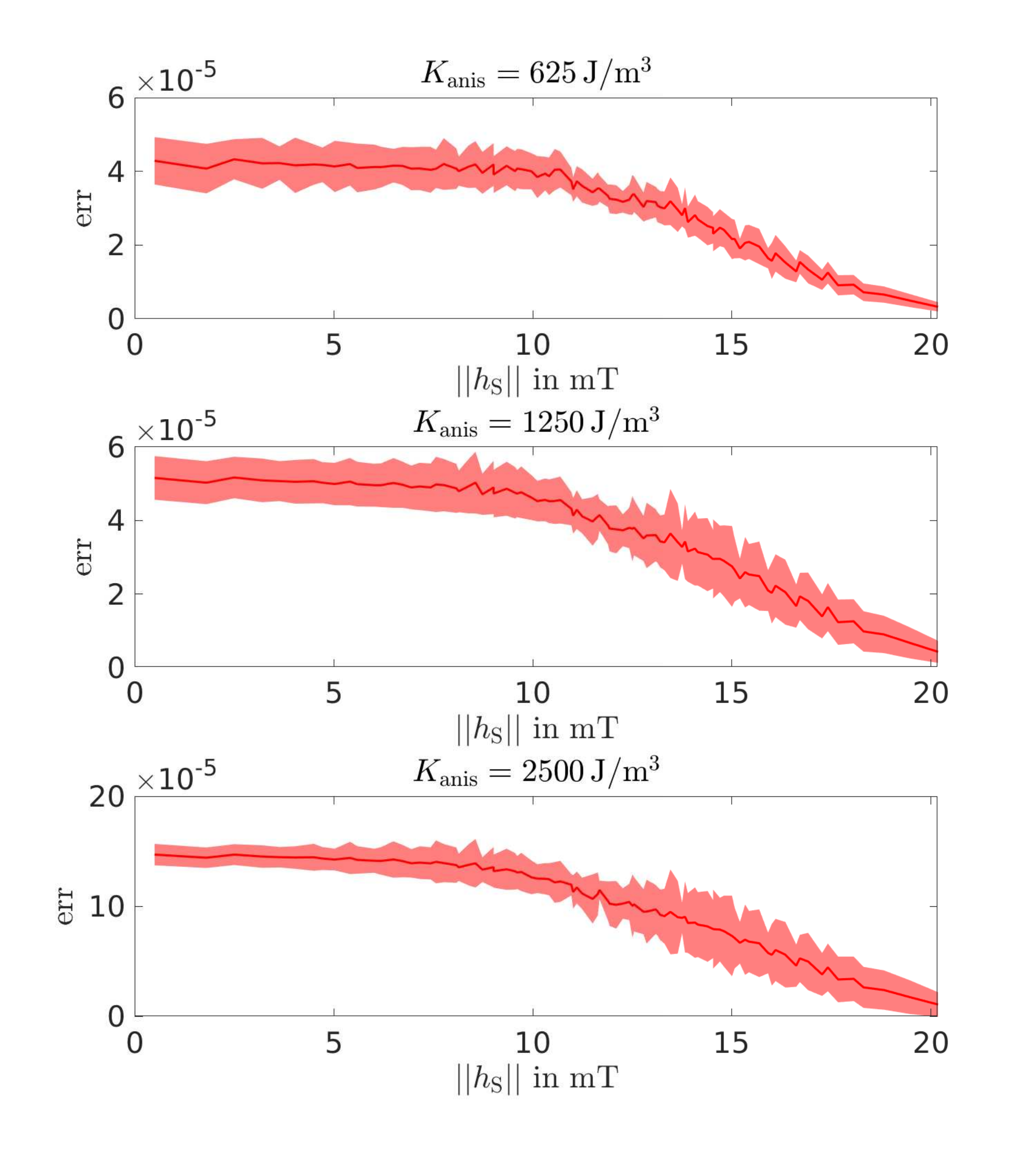}
    \end{minipage}
    \hfill
    \begin{minipage}{0.49\linewidth}
    \includegraphics[width=\linewidth]{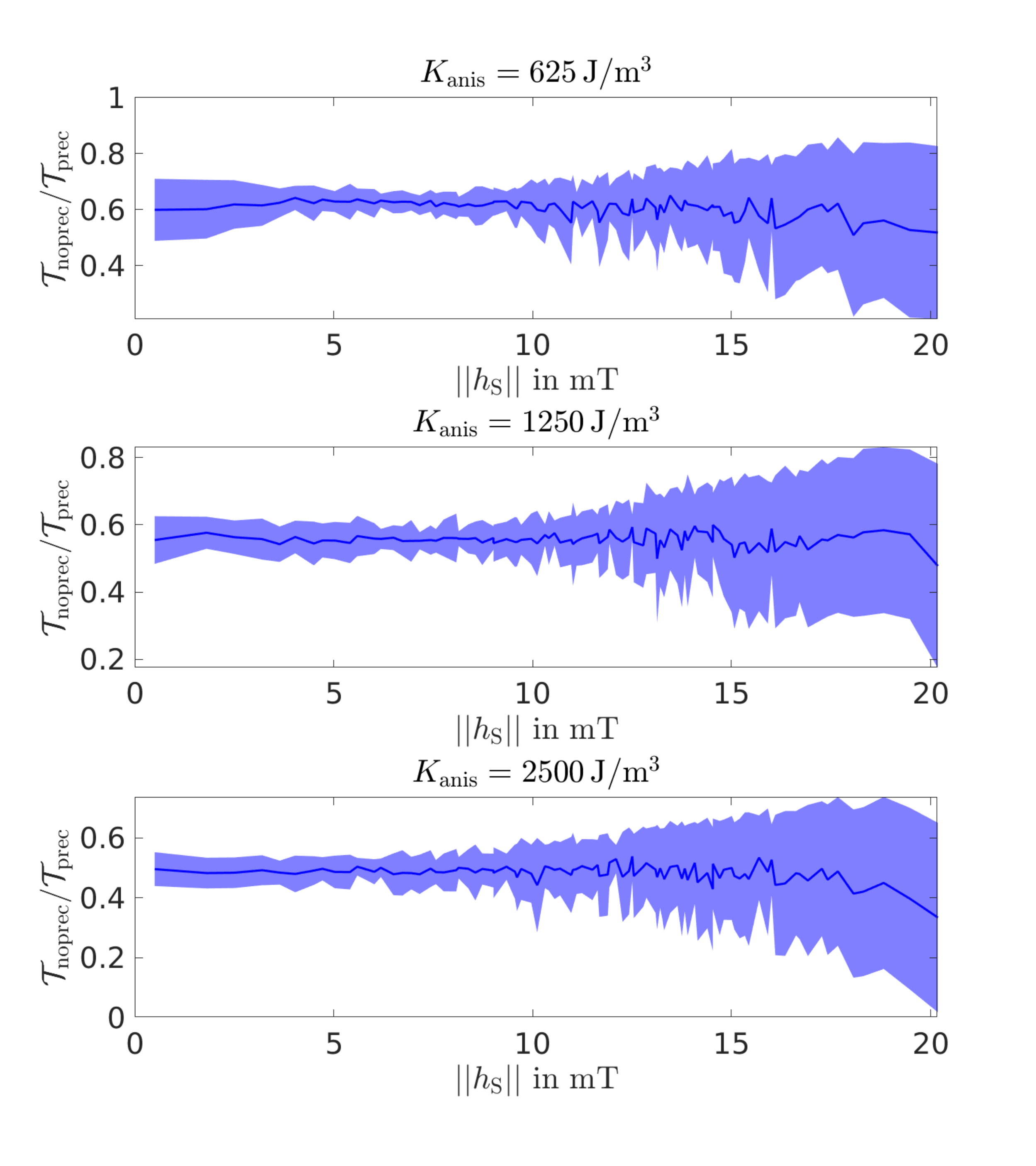}
    \end{minipage}
    \caption{Left: Relative errors in $\partial_t \mathbf{\bar{m}}$ when neglecting the precession term for different anisotropy constants. The mean and standard deviation corresponding to different easy axis directions are plotted. Right: Ratio of computation time $\mathcal{T}$ with and without precession term, again for different easy axes. Both are plotted against the drive field magnitude.}
    \label{fig:prec}
\end{figure}
To validate these anecdotal findings, we computed the N\'{e}el relaxation of MNPs for a typical 2D MPI scenario: We consider a field of view $\Omega$  of $30\times 30$ pixels (1 pixel corresponds to 1\,mm $\times$ 1\,mm; associated with positions $x_i\in\Omega$ ) with a static selection field $\mathbf{H}_\mathrm{S}:\Omega \rightarrow \R^3$ that is linear in space and a sinusoidal drive field $\mathbf{h}_\mathrm{D}(t)$. The applied field is then given as $\mathbf{H}(t)=\mathbf{h}_\mathrm{D}(t)+h_\mathrm{S}$ with $h_\mathrm{S} \in \{ \mathbf{H}_\mathrm{S}(x_i) | i=1, \hdots, 30^2 \}$ for each pixel, as described in Definition \ref{def:MPI-model-advection-diffusion}. This numerical experiment was conducted for a nanoparticle core diameter of $\SI{20}{nm}$, different easy axis angles and different anisotropy constants $K_\mathrm{anis}=625,1250,\SI{2500}{J/m^3}$ with and without the precession terms. The relative error as well as the computation times were recorded depending on the selection field magnitude $||h_S||$. Due to the linear structure of the selection field, low values of $||h_S||$ correspond to the center of the field of view, while high values correspond to its outer regions. The results are shown in Figure \ref{fig:prec}. Relative errors made when neglecting the precession terms are small, and get even smaller the larger the magnetic field is (i.e., at the edge of the field of view), but get larger as $K_\mathrm{anis}$ grows. Also, solving the PDEs without the precession terms usually takes a little over half as long as solving the full problem. However, at the edge of the field of view, some solves take less than a tenth the time. This improves computation times for the complete 2D region significantly -- up to a factor of four in our simulations.

Since the error in the particle signal due to this approximation is low while the gain in computation time is significant, we conclude that for most cases, neglecting the precession terms is advisable. Special consideration may be necessary for very large fields or anisotropy constants.
\subsection{Accuracy}
In order to estimate the computational accuracy of our methods, we simulated a one-dimensional excitation field for various parameters and compared the results. In particular, we set different core diameters from $30$ to $\SI{60}{nm}$ and different anisotropy constants from $400$ to $\SI{11000}{J/m^3}$. The particle response was then simulated using the spherical harmonics method, with $N_\mathrm{max}=20,\dots,60$ and the finite volume method with mesh level $3,\dots,6$. For the finite volume method, a triangulation of the sphere based on recursive partitioning of icosahedrons was used \cite{Gagarinov2017}. Mesh level 3 features 1280 spherical triangles discretizing the sphere, and for each mesh level this number gets quadrupled. Thus, mesh level 6 features $\SI{245760}{}$ spherical triangles.

\begin{figure}[h]
    \centering
    \includegraphics[width=\textwidth]{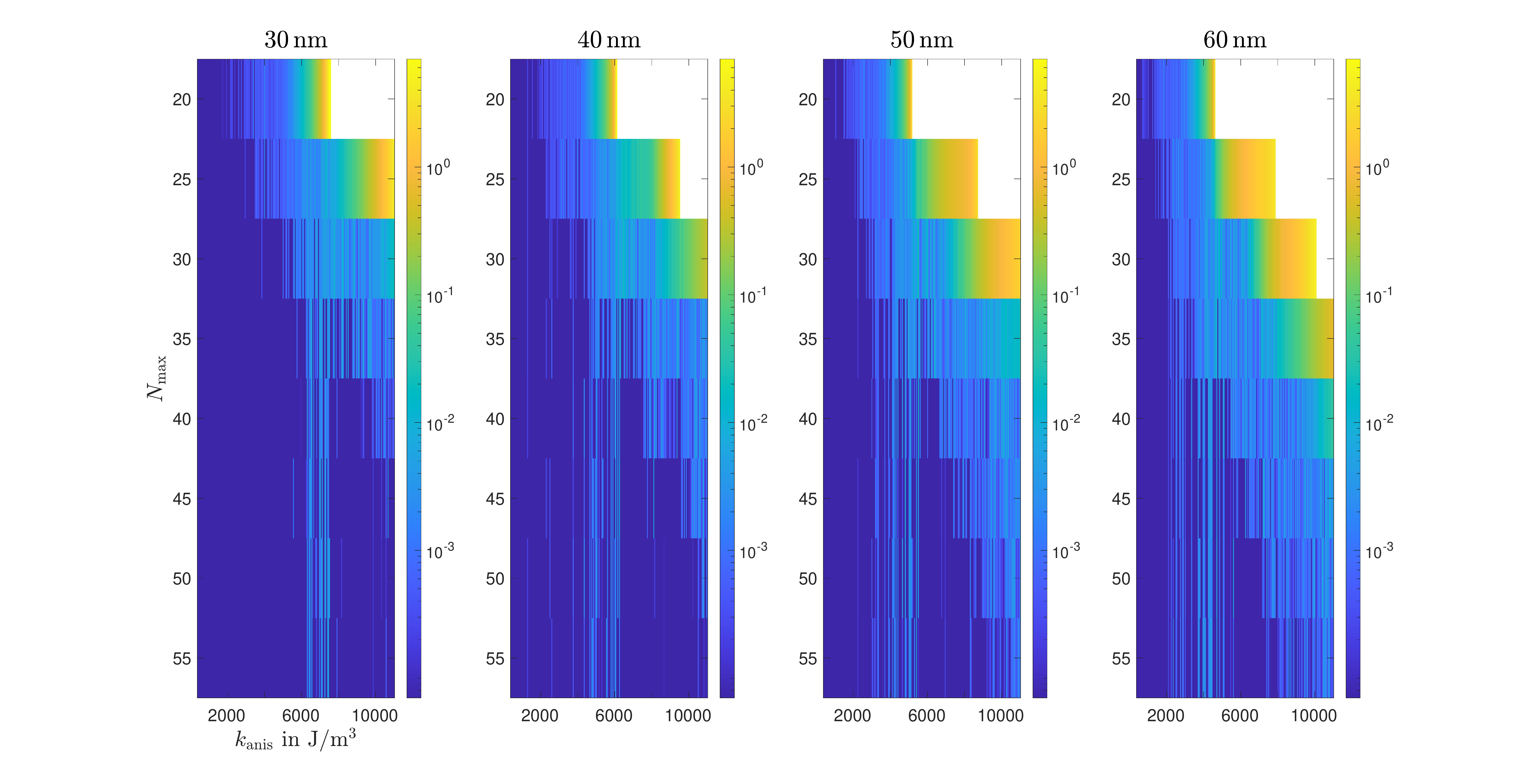}
    \caption{Relative error of the mean magnetic moment as obtained by the spherical harmonics method w.r.t. the $N_\mathrm{max}=60$ solution.}
    \label{fig:sh_acc}
\end{figure}

\begin{figure}[h]
    \centering
    \includegraphics[width=\textwidth]{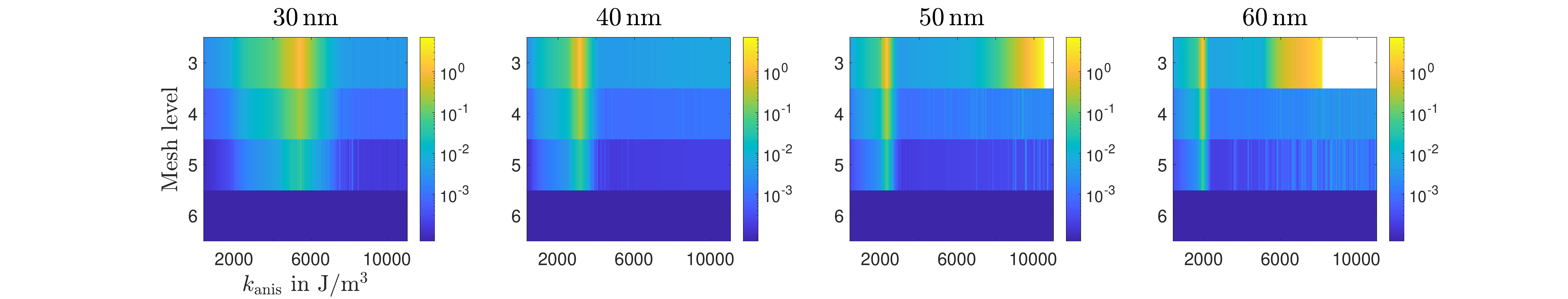}
    \caption{Relative error of the finite volume method results. Mesh level 6 was used as a reference.}
    \label{fig:fv_acc_beta0}
\end{figure}

The relative errors with respect to the finest discretization, i.e. mesh level 6 and $N_\mathrm{max} = 60$ respectively, are shown in Fig. \ref{fig:sh_acc} and \ref{fig:fv_acc_beta0}. When no color is shown, the solver was unable to solve the differential equation or the result was unphysical, i.e. the mean magnetic moment had a magnitude larger than the individual magnetic moments $m_0$.

\begin{figure}[h]
    \centering
    \includegraphics[width=\textwidth]{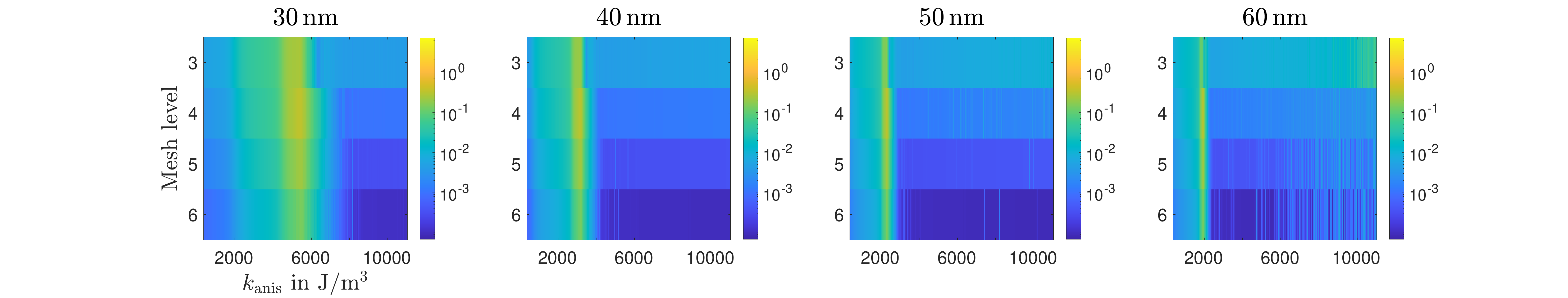}
    \caption{Relative FV error with 20\% upwind discretization. Mesh level 6 without upwind was used as a reference.}
    \label{fig:fv_acc_beta20}
\end{figure}

While the SH method requires a large maximum coefficient index $N_\mathrm{max}$ to even converge for large diameters and anisotropy constants, the FV method is more stable. However, the FV method exhibits a striking structure where the error is highest for a certain anisotropy constant and decreases to both sides for fixed particle diameter. This effect persists when the tolerances for the ODE solver are lowered. This makes it plausible that the effect stems from the spatial discretization itself. Introducing 20\% upwind discretization to the computation of the advection term at the triangle edges reduces this error somewhat, as shown in Fig. \ref{fig:fv_acc_beta20}.

\section{Selected parameter identification problems for the mean magnetic moment vector in the context of MPI} \label{sec:parameter_identification}

Modeling the system function $s$ in \eqref{eq:MPI_static_concentration} is one of the open key problems in MPI. 
It is strongly linked to the behavior of the mean magnetic moment $\mathbf{\bar{m}}$ of the nanoparticles in the applied magnetic field $\mathbf{H}$. Depending on environmental conditions and particle properties one can use an approximation by one dominating dynamic effect, i.e., Brown or N\'{e}el rotation. 
Given an appropriate model for the mean magnetic moment $\mathbf{\bar{m}}$ relying on parameters $\tilde{p}\in \tilde{\mathcal{P}}$, the calibration problem becomes a parameter identification problem with respect to $\tilde{p}$ for a given set of measured voltage-concentration tuples $(c^{(j)},v^{(j)})$.  

The parameter identification setups are outlined in the following (formulated for one receive coil unit): 
Using the Fokker-Planck equation in Section \ref{sec:magnetization_models} to describe the behavior for large ensembles of particles, here we represent the mean magnetic moment by the probability density function $f:\Omega \times S \times I \times \mathcal{P} \rightarrow \mathbb{R}_+$ depending on parameters $p$ in the finite-dimensional space $\mathcal{P}$.
Note that we included the spatial dependence of the offset fields in the field-of-view $\Omega$ explicitly, i.e., $\mathbf{H}:\Omega \times I \rightarrow \R^3$. 
\begin{itemize}
\item[E1] In the first example the computational model of the present work is used to generate a dictionary, which is able to explicitly model the variability of particle properties in polydisperse tracers, and which can be used to explain measured MPS data by reconstructing a weighting function. 
Varying parameters of individual particles can strongly influence the behavior of the tracer, i.e., one needs to take into account a distribution among certain parameters $p \in \mathcal{P}$ where these model parameters are assumed to be finite-dimensional. Given the specific mean magnetic moment $\mathbf{\bar{m}}_p$ for a selected parameter $p \in \mathcal{P}$ ($\mathcal{P}$ comprising a subset of the model parameters) we consider $\mathbf{\bar{m}}= \int_\mathcal{P} w(p) \mathbf{\bar{m}}_p \ \mathrm{d} p$ where $w$  is a nonnegative weighting function characterizing the tracer composition, i.e., $\tilde{p}=w$ and for example $\tilde{\mathcal{P}}\subset L^2(\mathcal{P})$.
Subsequently, we exploit a variational method to obtain the infinite-dimensional parameter $w$ for a given set of measured voltage-concentration tuples $(c^{(j)},v^{(j)})$. 
Precisely, the model for the mean magnetic moment is assumed to be given by 
\begin{equation}\label{eq:polydisperse_sumption_general}
    \mathbf{\bar{m}}(w)(x,t)= \int_{\mathcal{P}} w(p) \underset{=\mathbf{\bar{m}}_p (x,t)}{\underbrace{m_0 \int_{S^2} m f(x,m,t,p) \ \mathrm{d} m}} \mathrm{d}p.
\end{equation}

\item[E2] In the second example the computational model is used to simulate concentration-voltage tuples $(c^{(j)},v^{(j)})$ for certain model parameters $\tilde{p}\in\tilde{\mathcal{P}}$. The simulated measurements are then used to determine a convolution kernel in an approximate model which is used in the so-called $x$-space method \cite{goodwill2010x, Goodwill2011} for sinusodial as well as pulsed excitation patterns \cite{tay2019pulsed}.  
This approach relies on the general assumption that the mean magnetic moment vector is represented in terms of the trajectory of the so-called field free point, i.e.,
\begin{equation}\label{eq:kernel_assumption_general}
  \mathbf{\bar{m}}(\boldsymbol{\kappa})(x,t)= \int_0^t \boldsymbol{\kappa}(\mathbf{x}_\mathrm{FFP}(\tau) -x) \frac{\partial}{\partial \tau} \mathbf{x}_\mathrm{FFP}(\tau) \ \d \tau  + \text{const.}
\end{equation}
where $\boldsymbol{\kappa}:\R^3 \rightarrow \R^{3\times 3}$ is a matrix of convolution kernels and where $\mathbf{x}_\mathrm{FFP}:[0,T] \rightarrow \R^3$ is the trajectory of the field free point implicitly defined by $\mathbf{H}(\mathbf{x}_\mathrm{FFP}(t),t)=0$. The general formulation is motivated by the representation of the MPI forward operator in \cite[Def. 1]{Kluth2017} for the equilibrium model.  Note that the $x$-space method \cite{goodwill2010x, Goodwill2011} is a special case of \eqref{eq:kernel_assumption_general} and the subsequently defined observation operator for Cartesian excitation sequences (which is specified in more detail in the corresponding subsection). The fitting is then performed by determining the kernels in $\boldsymbol{\kappa}$, i.e., $\tilde{p}=\boldsymbol{\kappa}$.
\end{itemize}

In either case the link between the parameters $\tilde{p}$ in the mean magnetic moment and the measured concentration-voltage data tuples $(c^{(j)},v^{(j)})$ is given by $C^{(j)}(\mathbf{\bar{m}}) = v^{(j)} $ for the observation operator given by
\begin{equation}\label{eq:observation_operator}
C^{(j)}(\mathbf{\bar{m}}(\tilde{p}) )  =   a\ast \left(- \mu_0  \int_\Omega c^{(j)}(x) \mathbf{p}^R(x) \cdot \frac{\partial}{\partial t} \mathbf{\bar{m}}(\tilde{p})(x,\cdot) \mathrm{d} x\right)
\end{equation}
where $a$, $\mathbf{H}$, and $\mathbf{p}^\mathrm{R}$ are assumed to be given.

\subsection{Quantitative comparison with MPS measurements of immobilized particles (E1)}\label{MPS-section}

\begin{figure}
    \centering
    \includegraphics[width=0.8\textwidth]{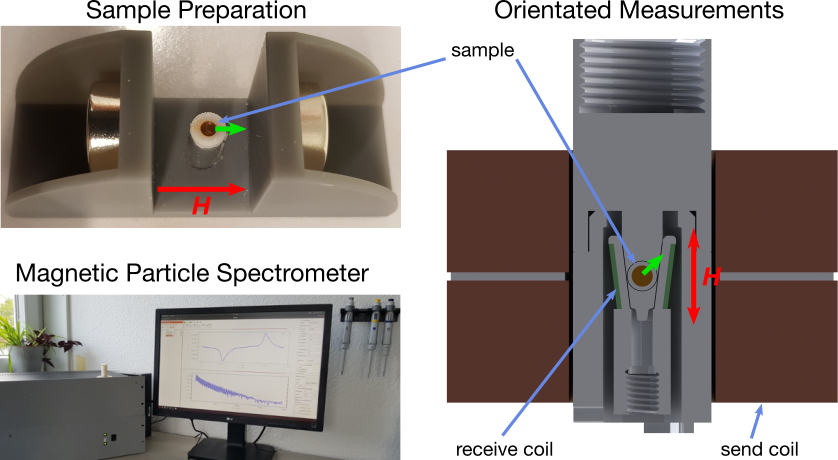}
    \caption{Overview of the experimental setup. The samples are prepared by aligning the easy axis of the particles using two neodymium magnets (upper left). The samples are immobilized using dental cement. A magnetic particle spectrometer (lower left) that applies a homogeneous oscillating magnetic field. The angle between the particle easy axis and the applied field is varied by manual rotation of the sample (right). The green arrow in the pictures indicates the orientation of the particles' easy axis.  }
    \label{fig:experimentalSetup}
\end{figure}

In order to show that the developed particle dynamic computation framework is capable of describing experiments, we  measured data with a custom magnetic particle spectrometer (MPS) that uses a sinusoidal 1D excitation with an excitation frequency of 125\,MHz$/4800\approx26041.66$\,Hz and an excitation amplitude of 20\,mT/$\mu_0$. The MPS uses a gradiometric receive concept \cite{Biederer2009d} and in turn also records the signal at the excitation frequency. The induced voltage signal is corrected for the transfer function of the receive path and thus contains the time derivative of the mean magnetic moment $\mathbf{\bar{m}}(t)$. A sample consisting of $V=13.5$ $\mu$L fluid particle solution (perimag, micromod GmbH, Germany) with a concentration $c_0=50$
~mmol L$^{-1}$ is immobilized in a disk-like sample holder using dental cement. During the preparation, two neodymium magnets are used to align the particles in a certain direction orthogonal to the rotation axis. The applied field during immobilization is 180~mT. The immobilized sample is mounted in a sample holder that allows to manually rotate the sample. In this way, three measurements were performed in the MPS at angles $ \phi_\mathrm{ref} \in \{0\,^\circ,45\,^\circ, 90\,^\circ\}$ where $0\,^\circ$ means that the easy axis of the sample is fully aligned with the excitation field and $90\,^\circ$ means that the easy axis is orthogonal to the excitation direction.  See Fig.\,\ref{fig:experimentalSetup}, for the experimental setup.

Due to the characteristics of the tracer and the preparation of the samples, in  \eqref{eq:polydisperse_sumption_general} we assume a distribution $w$  with respect the particle diameter $D$, the anisotropy constant $K_\mathrm{anis}$, and the orientation $\phi$ of the easy axis $\mathbf{n}(\phi)=(\cos (\phi), \sin (\phi), 0)^t$ in the $x$-$y$-plane.
As we deal with an immobilized nanoparticle sample whose orientation is changed between experiments, $w$ does not depend on $\phi$ directly. 
We thus consider the distribution function $w$ with respect to $\Delta \phi = \phi - \phi_\mathrm{ref}$ as $\Delta \phi$ is now invariant with respect to a change of the orientation of the entire sample.  

To derive the specific problem setup we apply the following assumptions to the observation operator in \eqref{eq:observation_operator}. In the previously described MPS setup we have a receive coil unit which is oriented in $x$-direction and its sensitivity profile is parallel to the unit vector in $x$-direction in good approximation, i.e., $\mathbf{p}^R (x) = k e_1$ for one constant $k>0$. Furthermore, the voltage data is corrected for the transfer function of the analog filter such that it can be neglected. In addition, no space-dependent selection field and no offset fields are applied (strictly speaking, a small effect of the earth's magnetic field can be observed). Together with the assumption that the drive field is homogeneous at the support of the homogeneous concentration function of the samples $c^{(j)}$, the probability density function $f$ is evaluated at $x=0$. In total there is no space dependence in the observation operator anymore. 
As a result, we consider the following linear inverse problem with respect to $w$ 
\begin{align}
v^{(j)}&=A^{(j)}(w) =  \int_{\mathcal{P}} w(D,K_\mathrm{anis},\underset{=\phi-\phi_\mathrm{ref}^{(j)}}{\underbrace{\Delta \phi}}) \Psi(D,K_\mathrm{anis},\phi_\mathrm{ref}^{(j)}+\Delta \phi)  \mathrm{d}(D,K_\mathrm{anis}, \Delta \phi) \quad, j=1,2,3. \label{eq:polydisperse_operator} 
\end{align}
with 
\begin{equation}
    \Psi(D,K_\mathrm{anis},\phi)= - \mu_0  c_0 V k m_0(D)   \int_{S^2}m_1  \frac{\partial}{\partial t} f(0,m,t,(D,K_\mathrm{anis},\mathbf{n}( \phi))) \  \mathrm{d}m   
\end{equation}
%
The weighting function $w$ is then obtained by minimizing the following Tikhonov-type functional
\begin{equation}\label{eq:mps_functional}
    J_\beta(w)= \sum_j \| A^{(j)}(w)- v^{(j)}\|_{L^2(I)}^2 + \beta \mathcal{R}(w)
\end{equation}
where the penalty term $\mathcal{R}$ is chosen as an $\ell^p$-norm (see for example \cite{jin2012sparsity}), which guarantees a stable solution to the problem. We further incorporate a positivity constraint and a symmetry assumption with respect to $\Delta \phi$, i.e., we include the constraint $w(D,K_\mathrm{anis},\Delta \phi)= w(D,K_\mathrm{anis},-\Delta \phi)$ in \eqref{eq:polydisperse_operator}.
Thus,
\begin{align}
A^{(j)}(w) =  \int_{\mathcal{P}_+} w(D,K_\mathrm{anis},\Delta \phi) \left( \Psi(D,K_\mathrm{anis},\phi_\mathrm{ref}^{(j)}+\Delta \phi) + \Psi(D,K_\mathrm{anis},\phi_\mathrm{ref}^{(j)}-\Delta \phi) \right)  \mathrm{d}(D,K_\mathrm{anis}, \Delta \phi)   
\end{align}
for $j=1,2,3$, with $\mathcal{P}_+=\mathcal{P}\cap (\R \times \R \times \R_+)$.

In order to approximate the distribution vector $w$, we first calculate $\Psi$ on a finite subset $\hat{\mathcal{P}} \subset \mathcal{P}$. In particular, we choose $\hat{\mathcal{P}} = \mathcal{P}_D \times \mathcal{P}_{K_\mathrm{anis}} \times \mathcal{P}_{\Delta \phi}$ with
\begin{align}
    \mathcal{P}_D &= \{16, 18, \dots, 58\}\SI{}{nm},\\
    \mathcal{P}_{K_\mathrm{anis}} &= \{450,500,\dots,6000\} \SI{}{J/m^3},\\
    \mathcal{P}_{\Delta \phi} &= \{-20,-15,...,+20\}^\circ.
\end{align}

The simulations, restricted to a fixed equidistant time grid, then represent the columns of a matrix $\hat{A}$ where $\hat{A}w$ approximates $A^{(j)}(w)$ as in \eqref{eq:mps_functional}. For the regularization functional $\mathcal{R}$, the $\ell^1$-norm was chosen, i.e., $\mathcal{R}(w) = \|w\|_1$. This prioritizes sparse solutions, i.e., solutions for which most of the entries of $w$ are zero. This is justified since the choice of $\hat{\mathcal{P}}$ incorporated no prior knowledge about the data, suggesting that most of the simulated parameters are not present in a real-world measurement. Also, on simulated data, the $\ell^1$-penalty term yielded better results compared to $\ell^2$-penalties. The simulated data was also used to determine the regularization parameter $\beta$.\\
\begin{figure}[h]
    \centering
    \begin{minipage}{0.49\linewidth}
    \includegraphics[width=\linewidth]{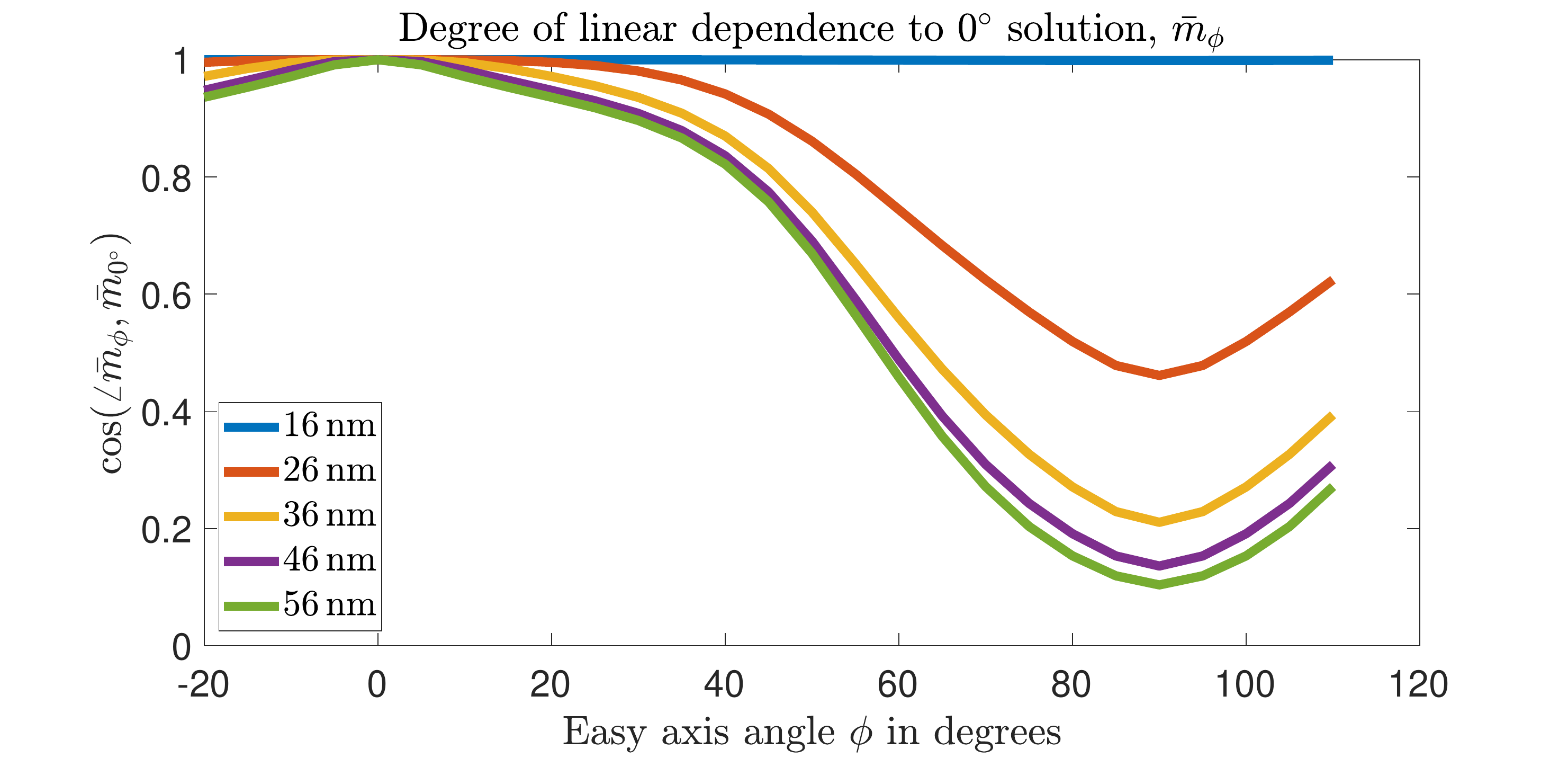}
    \end{minipage}
    \hfill
    \begin{minipage}{0.49\linewidth}
    \includegraphics[width=\linewidth]{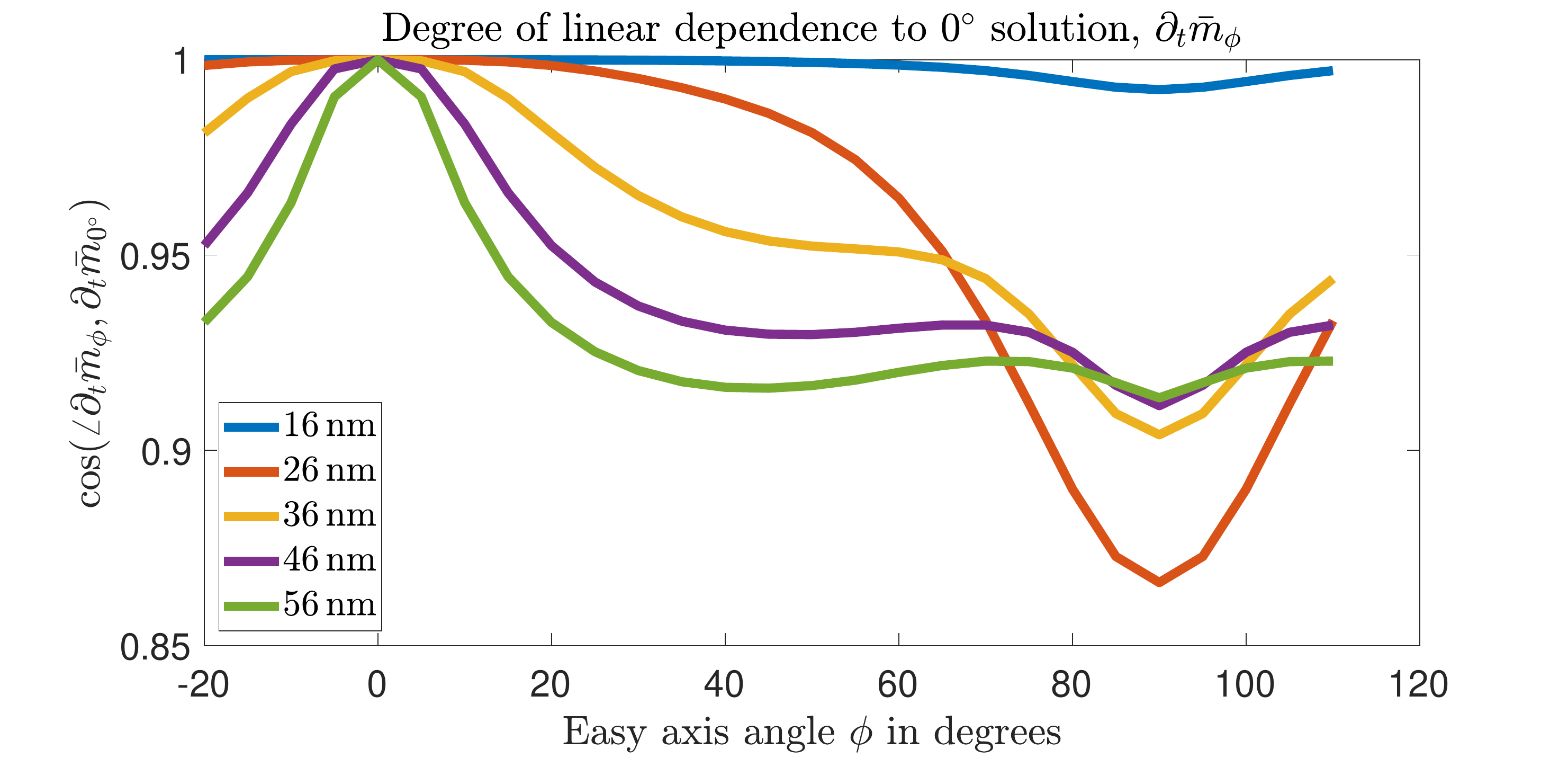}
    \end{minipage}
    \caption{Degree of linear dependence of simulated solutions for different easy axis angles relative to the $0^\circ$ solution for different core diameters, averaged over all anisotropy constants. The cosine of the angles between the respective solutions is displayed, for the expected magnetic moment as well as its derivative. It becomes clear that for small angle deviations, the dynamics of the nanoparticles are mostly scaled w.r.t. the reference angle.}
    \label{fig:mps_angles}
\end{figure}
After having computed the simulations, we noticed that small angle deviations lead to nearly no new information about the nanoparticle dynamics (Fig. \ref{fig:mps_angles}). The reason for this is likely the discarded information from the $y$-channel. Since the measured data is purely one-dimensional, we thus decided to omit the dependence on $\Delta \phi$ from the optimization, i.e., using $\mathcal{P}_{\Delta \phi} = \{0\}^\circ$, and leave more systematic investigations of determining orientation distributions for future works.

The minimizer of \eqref{eq:mps_functional} for a given regularization parameter $\beta$ is obtained by the nonnegative LASSO method as described in \cite{nnlasso}, where the restriction $w\geq 0$ was chosen in order to be able to interpret the entries of $w$ as weights applied to the simulated behavior of each parameter combination.

\begin{figure}[h]
    \centering
    \begin{minipage}{0.49\linewidth}
    \includegraphics[width=\linewidth]{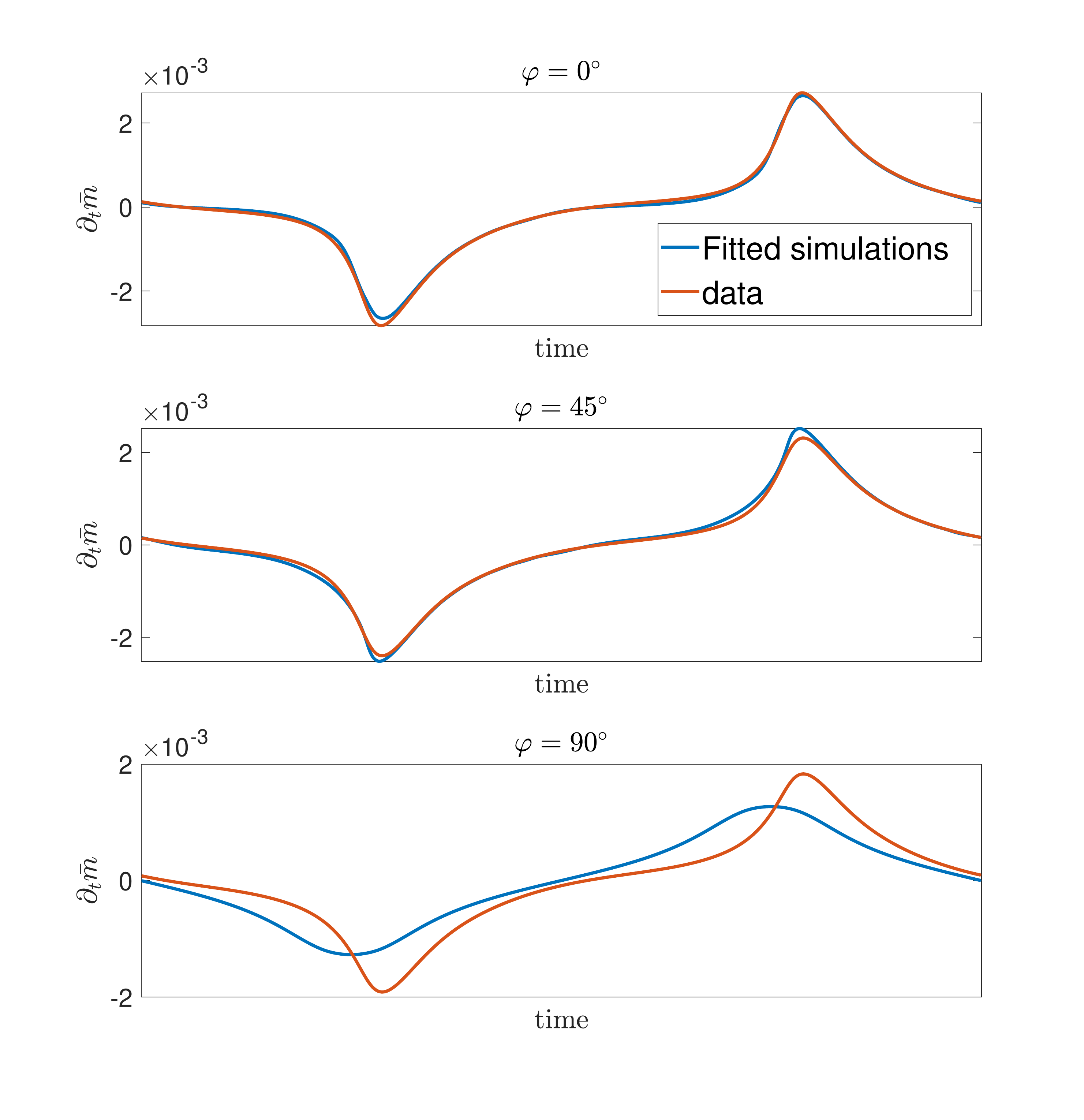}
    \end{minipage}
    \hfill
    \begin{minipage}{0.49\linewidth}
    \includegraphics[width=\linewidth]{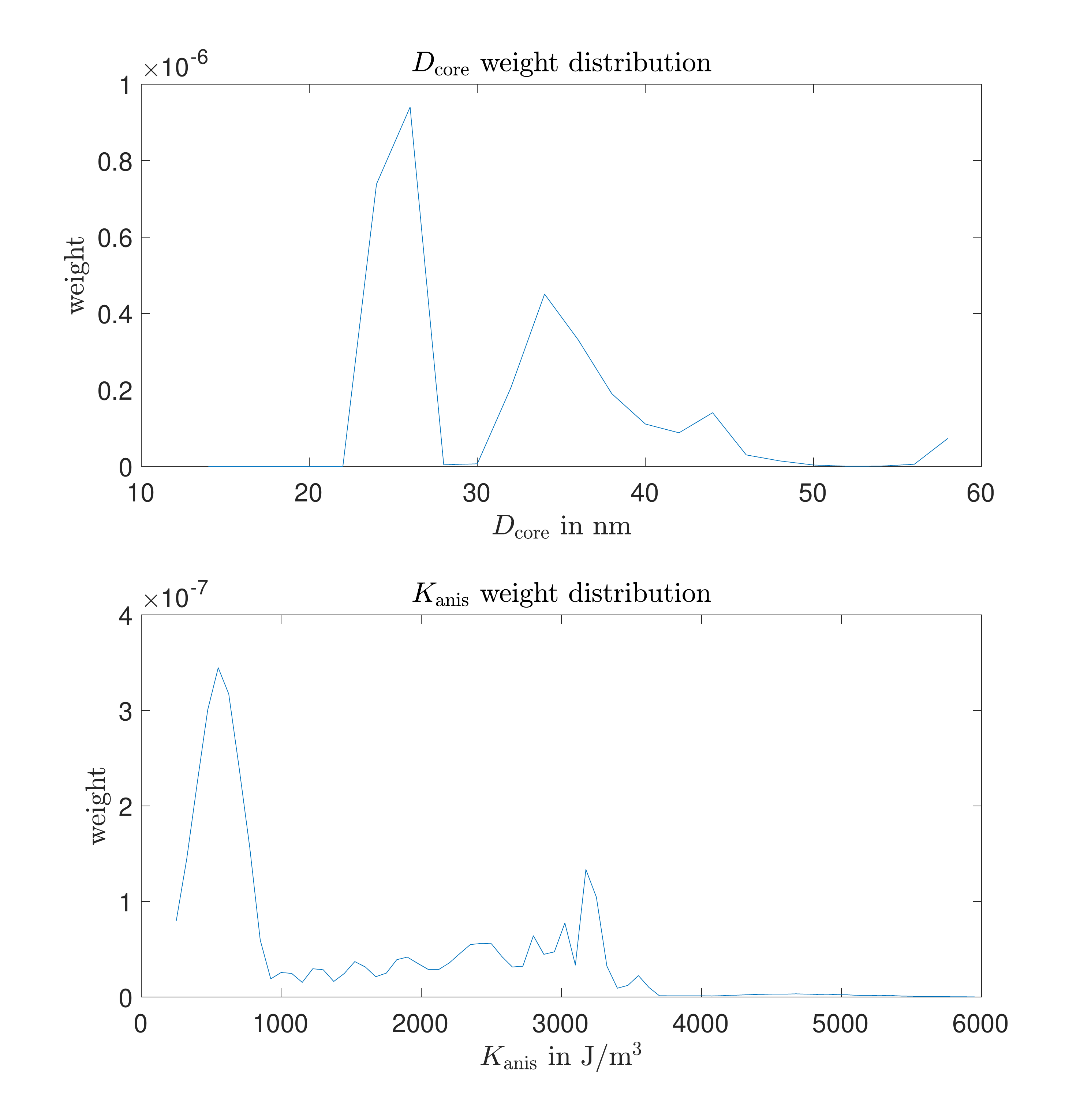}
    \end{minipage}
    \caption{Result of the optimization procedure described in section \ref{MPS-section}. On the left, it is shown that the simulations can explain the measured data for $\varphi=0^\circ$ and $45\,^\circ$, but not sufficiently for $\varphi = 90^\circ$. On the right, the marginal distributions of the weights with respect to the core diameter and the anisotropy constant are illustrated. This can be interpreted physically as a parameter distribution for the tracer material that approximately yields the measured data.}
    \label{fig:mps_nophi}
\end{figure}

The results of the procedure are shown in Fig. \ref{fig:mps_nophi}. It becomes apparent that the weighted simulations agree well with the measured data, except for the $90^\circ$ case, where the nanoparticles' easy axes were oriented perpendicularly to the applied field. The corresponding marginal distributions of the weights with respect to core diameter and anisotropy constant may be interpreted as a possible distribution of these parameters in the measured sample. However, since there are many different viable solutions yielding a good fit to data, we do not claim that this distribution reflects the physical reality.

\begin{figure}[h]
    \centering
    \begin{minipage}{0.49\linewidth}
    \includegraphics[width=\linewidth]{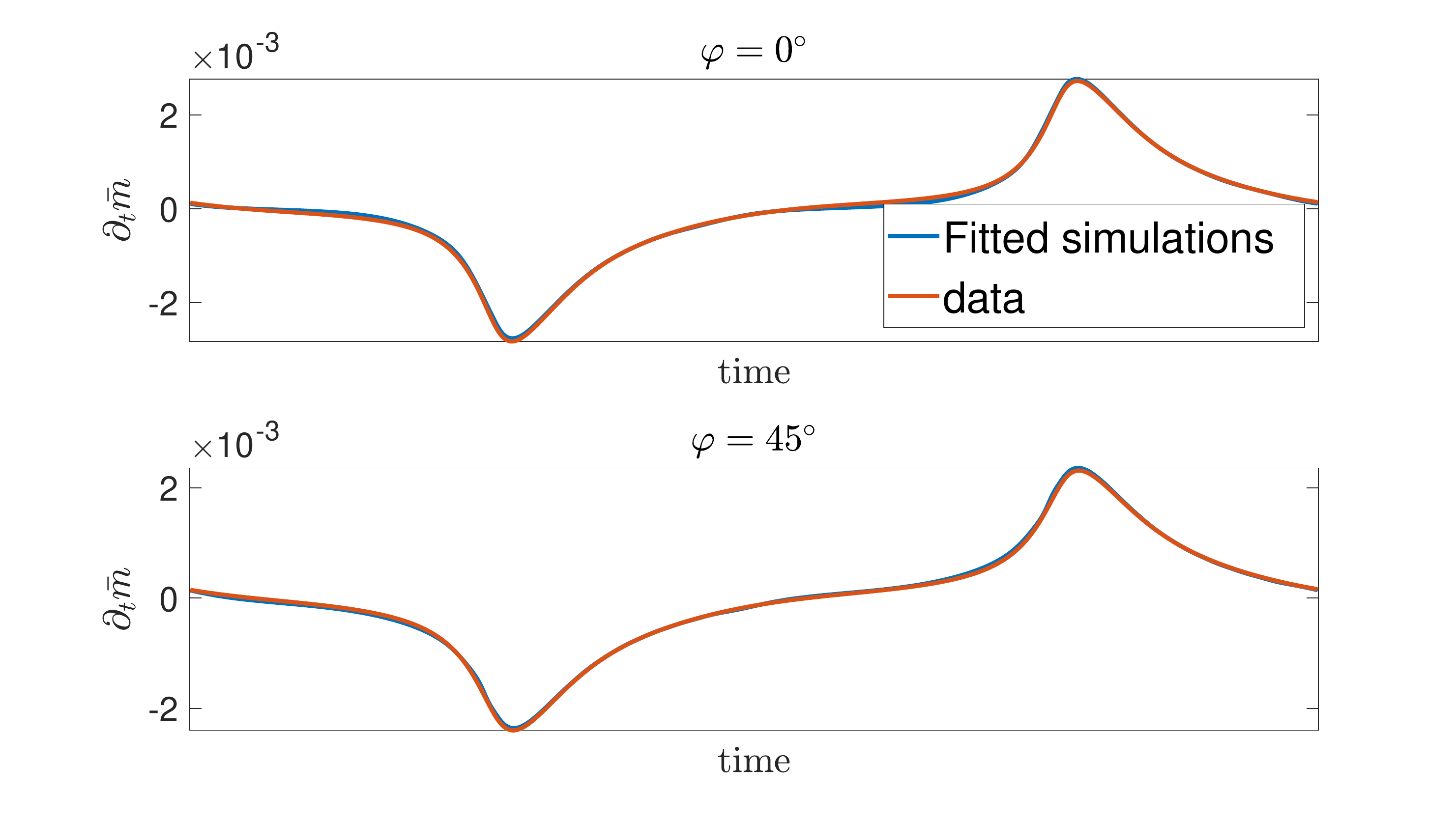}
    \end{minipage}
    \hfill
    \begin{minipage}{0.49\linewidth}
    \includegraphics[width=\linewidth]{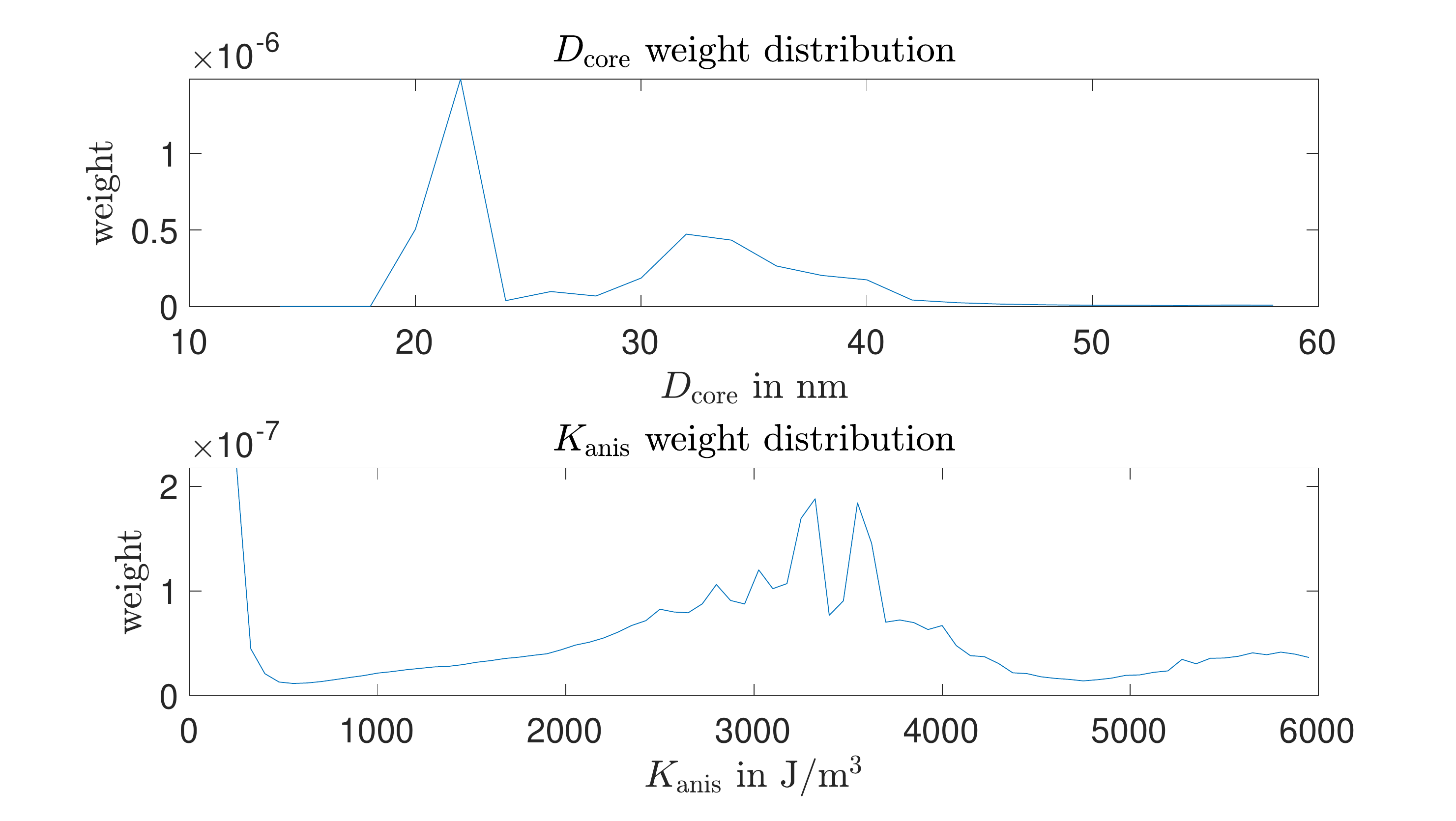}
    \end{minipage}
    \caption{Result of the same optimization procedure, where the measurements for $\varphi = 90^\circ$ were not included. The fit to the data is near perfect, and the weight distribution appear reasonable. However, caution has to be exerted when interpreting these results.}
    \label{fig:mps_nophi_no90}
\end{figure}

Since the simulations fail to describe the $90^\circ$ case, the calculation was also carried through for only $0^\circ$ and $45^\circ$ (Fig. \ref{fig:mps_nophi_no90}). The agreement with data is nearly perfect, and the marginal distributions are qualitatively similar to the results including $90^\circ$.
\subsection{Qualitative comparison with Cartesian and pulsed excitation results (E2)}
From the very general formulation in \eqref{eq:kernel_assumption_general} we derive two one-dimensional problem setups for sinusoidal and pulsed excitation \cite{tay2019pulsed}, which are commonly considered in the $x$-space method \cite{goodwill2010x, Goodwill2011}.

\paragraph{First we derive the sinusoidal setup of the $x$-space method} 
We assume a homogeneous sinusoidal excitation in $e_1$-direction ($e_i$, $i=1,2,3$, denote the unit vectors in $\R^3$), i.e., the drive field is given by $\mathbf{H}_\mathrm{D}(t)=A \sin (2 \pi f t) e_1$ with amplitude $A>0$ and excitation frequency $f>0$. The selection field is a approximately linear gradient field, i.e., $\mathbf{H}_\mathrm{S}(x)= Q_G x$ with $Q_G=G \mathrm{diag} (-0.5,-0.5,1)$ and $G>0$ being the gradient strength. We further neglect the analog filter and assume that the receive coil is sensitive to the $e_1$-direction and homogenenous in the region of interest, i.e., without loss of generality we assume $\mathbf{p}^R(x)=e_1$ for any $x\in \R^3$.

We then can define the field free point trajectory by $\mathbf{x}_\mathrm{FFP}(t)=-Q_G^{-1}\mathbf{H}_D(t)=\underset{=:x_\mathrm{FFP}(t)}{\underbrace{2 A/G \sin (2 \pi f t)}} e_1$ and measurement time $T=1/f$.
To derive the $x$-space method we make the rather artificial assumption that the support of the concentration function is restricted to the $e_1$-axis and that $c^{(j)}$ behaves like a delta distribution with regard to $e_2$ and $e_3$ directions. Technical details of the lower-dimensional problem derivation can be found in \cite[Sec. 2.2]{Kluth2018b}.
 Plugging all these assumptions into the observation operator \eqref{eq:observation_operator} together with the model assumption \eqref{eq:kernel_assumption_general}, we derive the following relation
 \begin{equation}\label{eq:pi_operator1}
     A^{(j)}(\kappa)\propto  \int_{\Omega_1} c^{j}(x) \underset{=:\kappa(x_\mathrm{FFP}(t)-x)}{\underbrace{\kappa_{1,1}(x_\mathrm{FFP}(t)-x,0,0)}}  \frac{\partial}{\partial t} x_\mathrm{FFP}(t) \mathrm{d} t,
 \end{equation}
 which maps the convolution kernel $\kappa: \R \rightarrow \R$ to the measurement. 
 Nevertheless, the $x$-space methods further assumes that data is considered on a subset $\tilde{I} \subset [0,T]$, where $x_\mathrm{FFP}$ is bijective and $\frac{\partial}{\partial t} x_\mathrm{FFP}(t)\neq 0$ for any $t\in \tilde{I}$.
 Then, the problem equation is adapted such that 
 \begin{equation}
     g^{(j)}(\bar{x}):=\frac{v^{(j)}(x_\mathrm{FFP}^{-1}(\bar{x}))}{\frac{\partial}{\partial t} x_\mathrm{FFP}(x_\mathrm{FFP}^{-1}(\bar{x}))} = B^{(j)}(\kappa)(\bar{x}):= (\kappa \ast c^{(j)})(\bar{x}).
 \end{equation}
 The previous assumptions are exploited to reformulate the integral as a convolution implying restrictions on the available information as $\bar{x}\in x_\mathrm{FFP}(\tilde{I}) \subset (-\frac{2A}{G},\frac{2A}{G})$. 
 This also relates to the denominator, for which holds $\frac{\partial}{\partial t} x_\mathrm{FFP}(x_\mathrm{FFP}^{-1}(\bar{x}))=2 \pi f \sqrt{(\frac{2A}{G})^2 - \bar{x}^2}$.
 The formulation as a convolution also requires some restrictions on the support of either $c^{(j)}$ ($\mathrm{supp} \ c \subset \Omega_1$) or $\kappa$ ($\mathrm{supp} \ \kappa \subset (- \frac{2A}{G} - \sup \Omega_1, \frac{2A}{G} - \inf \Omega_1)$). Note that the assumption on the support of the concentration has already been exploited to restrict the integral to $\Omega_1$. $g^{(j)}$ comprises FFP speed normalization and gridding.
 
 Given some phantom voltage tuples $(c^{(j)}, v^{(j)})$, $j=1,\hdots,N_J$, the kernel estimation problem is then solved by minimizing the functional 
 \begin{equation}
     J(\kappa) = \sum_j \| c^{(j)} \ast \kappa - g^{(j)} \|_{L^2(\R)}^2 = \sum_j \| \mathcal{F}(c^{(j)}) \mathcal{F}(\kappa) - \mathcal{F}(g^{(j)}) \|_{L^2(\R)}^2,
 \end{equation}
where we  use the operator $B^{(j)}$, obtain $g^{(j)}$ from $v^{(j)}$ as previously described, and where $\mathcal{F}$ is the Fourier transform. 
 The final solution is then obtained via discretization of the following representation of the minimizer of $J$
 \begin{equation}\label{eq:xspace_calib}
     \kappa = \mathcal{F}^{-1} \left ( \left(\sum_j \overline{\mathcal{F}(c^{(j)})} \mathcal{F}(c^{(j)}) \right)^{-1} \sum_j \overline{\mathcal{F}(c^{(j)})} \mathcal{F}(g^{(j)}) \right),
 \end{equation}
 which is obtained from first order optimality conditions of $J$.
 Using equidistant nodes $\{x_i\}_{i=1,\hdots,N} \subset (-\frac{2A}{G},\frac{2A}{G})$ we obtain the discretized solution $\tilde{\kappa}\in \R^N$ ($\tilde{c}^{(j)} \in \R^N$ being the discretized concentration and $\tilde{g}^{(j)} \in \R^N$ the discretized processed measurement) by
 \begin{equation}\label{eq:kernel_extimate_sin}
     \hat{\tilde{\kappa}}_k = \frac{\sum_j \overline{\hat{\tilde{c}}^{(j)}_k} {\hat{\tilde{g}}^{(j)}_k}}{\sum_j \overline{\hat{\tilde{c}}^{(j)}_k} {\hat{\tilde{c}}^{(j)}_k}}. 
 \end{equation}
 The stability of the solution strongly relies on the concentration phantoms $c^{(j)}$, which are used for the purpose of calibration.
 Depending on their choice the functional $J$ might be extended to include some kind of a priori knowledge to obtain a regularized and stable solution to the problem. 
 In the discretized setting we use delta samples which guarantee a stable solution. 
 From a theoretical point of view, one single calibration measurement would then be sufficient to identify the point spread function/convolution kernel (under the assumption that the system behaves linearly and is shift invariant with respect to the FFP position).
 However, we can only expect that the measurements $v^{(j)}$ obtained by the simulation toolbox fulfill the shift invariance in good approximation.  
 
 \begin{remark}
 The $x$-space image reconstruction can be performed analogously to \eqref{eq:xspace_calib} for $N=1$ (by changing the roles of $\kappa$ and $c^{(1)}$) when for example determining a concentration $c^{(1)}$ for given $\kappa$ and gridded speed normalized phantom measurement $g^{(1)}$. Depending on the structure of $\kappa$, i.e., if it is sufficiently close to a $\delta$ distribution, the $g^{(1)}$ can be used as reconstruction directly, which is also done in practice \cite{goodwill2010x, Goodwill2011}. Numerically and experimentally this can be realized by suitable combinations of gradient strength and spatial discretization. If $\kappa$ does not have this specific structure, one needs to solve the deconvolution problem. A stable solution is then not given analogously to \eqref{eq:xspace_calib} and additional a priori information on the concentration needs to be included in the reconstruction method.   
 \end{remark}

 \begin{remark}
 Note that the speed normalization and the gridding introduce an implicit weighting when compared to the original measurement. This can be disadvantageous for a proper noise treatment when using real measurements and when aiming for a variational approach to obtain the kernel. As an alternative one could consider the parameter identification problem using the operator $A^{(j)}$ in \eqref{eq:pi_operator1} and a corresponding functional 
 \begin{equation*}
 J(\kappa) = \sum_j \| A^{(j)} \kappa - v^{(j)} \|_{L^2([0,T])}^2
 \end{equation*}
for noise treatment in the measurements $v^{(j)}$. 
Further investigations in this direction are beyond the scope of this work and remain future work.
 \end{remark}

 \paragraph{Second, we derive the setup of the pulsed MPI method} While the previous method has a longer history, the pulsed sequence approach has been recently proposed. The authors in \cite{tay2019pulsed} derived the model equation from a physical point of view. In contrast, we derive the pulsed sequence approach starting with the model \eqref{eq:kernel_assumption_general}.
 Again, we restrict the problem to obtain the concentration with respect to the $e_1$-direction, i.e., on $\Omega_1$ as previously defined.
 In contrast to the sinusoidal excitation, we have two applied field dynamics included to obtain the desirable information. 
 A pulsed fast excitation is applied in an orthogonal direction to $e_1$, e.g., without loss of generality in the $e_2$-direction. This is accompanied by a slower step-wisely changing field in $e_1$-direction. 
 The pulsed excitation causes the main change of the nanoparticles' magnetization while the second one guarantees the spatial encoding along the $e_1$-direction (i.e, information is obtained in a frame-by-frame manner).
 More precisely in the terminology of the previous setup, we have a different drive field 
 \begin{equation}
     \mathbf{H}_\mathrm{D} (t) = A \phi_{[-1,1],T_\text{pulsed},N}(t) e_1 + A_\mathrm{pulsed} \text{sign}(\sin(2 \pi f_\mathrm{pulsed} t) e_2 
 \end{equation}
 with $f_\text{pulsed}=1/T_\text{pulsed}$ and with 
 \begin{equation}
     \phi_{[a,b],T_\text{pulsed},N}(t)=\frac{b-a}{N-1} \left\lfloor \frac{t-\Delta t}{T_\text{pulsed}} \right\rfloor + a,
 \end{equation}
 where $\Delta t$ is a small time shift which ensures that changes in the different directions do not appear simultaneously. 
 Here, the choice of $N$ determines the spatial discretization in $e_1$-direction as becomes clear below.
 We thus obtain 
 \begin{equation}
     \mathbf{x}_\mathrm{FFP} (t) = - Q_G^{-1} \mathbf{H}_\mathrm{D} (t) = \underset{=: x_\mathrm{FFP}(t)}{\underbrace{\frac{2A}{G} \phi_{[-1,1],T_\text{pulsed},N}(t) }} e_1 + \underset{=: y_\mathrm{FFP}(t)}{\underbrace{\frac{2 A_\text{pulsed}}{G} \text{sign}(\sin(2 \pi f_\mathrm{pulsed} t) }} e_2.
 \end{equation}
Besides the different composition of the drive field, the pulsed sequence approach also exploits the signal from a different receive coil unit. Information about the concentration function is derived from a receive coil being sensitive to the $e_2$-direction, i.e., $\mathbf{p}^R(x)=e_2$.
Analogously to the sinusoidal excitation approach we start from the general model \eqref{eq:kernel_assumption_general}. 
Plugging in the previous assumptions in the observation operator \eqref{eq:observation_operator} we derive the relationship 
\begin{equation}\label{eq:pulsed_voltage}
    v^{(j)}(t)= \int_{\Omega_1} c^{(j)} ( \kappa_{2,2} ( x_\mathrm{FFP}(t) -x, y_\mathrm{FFP}(t),0) \frac{\partial}{\partial t} y_\mathrm{FFP}(t) +  \kappa_{2,1} ( x_\mathrm{FFP}(t) -x, y_\mathrm{FFP}(t),0) \frac{\partial}{\partial t} x_\mathrm{FFP}(t) ) \ \mathrm{d} x.
\end{equation}
Note that this equation needs to be interpreted in the sense of distributions as the time derivatives of $x_\mathrm{FFP},y_\mathrm{FFP}$ do not exist in the classical sense. We thus define a distribution (linear and continuous mapping) which yields a real number when applied to the previous term. 
We compute the integral over certain time intervals $I_i\subset I$ which are characterized by two important properties: First, they do not include the change of $x_\mathrm{FFP}$ but they include the change of $y_\mathrm{FFP}$. And second, the length of the time interval is sufficiently large such that the system reaches a steady state again.
We thus determine real numbers $\tilde{g}_i^{(j)}$ from the measurement signal $v^{(j)}(t)$ in \eqref{eq:pulsed_voltage} as follows
\begin{align}
    \tilde{g}_i^{(j)}= (-1)^i \int_{I_i} v^{(j)}(t) \ \mathrm{d} t & = (-1)^i \int_{\Omega_1} c^{(j)}(x) \int_{I_i} \kappa_{2,2}(x_\mathrm{FFP}(t) - x , y_\mathrm{FFP}(t), 0) \frac{\partial}{\partial t} y_\mathrm{FFP}(t) \ \mathrm{d} t \mathrm{d} x \notag \\
    & =   \int_{\Omega_1} c^{(j)}(x)  \underset{=:\kappa(x_i-x)}{\underbrace{\kappa_{2,2}(x_i - x , 0, 0) } \mathrm{d} x} \notag \\
    &=  (c^{(j)}\ast \kappa ) ( x_i) ,
\end{align}
where the alternating sign results from the construction of $y_\mathrm{FFP}$.
We thus have obtained an analogous problem setup compared to the sinusoidal case, which is based on the pulsed sequence and an alternative processing of the measured data $v$.

\begin{remark}
 The distributional interpretation is a rather artificial point of view when starting from the general model assumption \eqref{eq:kernel_assumption_general}. One may choose an alternative starting point such that one can consider Dirac sequences for $\frac{\partial}{\partial t}x_\mathrm{FFP}$ and $\frac{\partial}{\partial t} y_\mathrm{FFP}$ which would converge to the previous setup. From a physical point of view this would be reasonable as a perfect jump discontinuity would not be realizable in practice. 
\end{remark}

The kernel $\kappa$ is then determined analogously by minimizing the functional 
\begin{equation}
    J_\text{pulsed}(\kappa) = \sum_j \| ((c^{(j)} \ast \kappa)(x_i))_{i=1,\hdots,N} - \tilde{g}^{(j)} \|_2^2.
\end{equation}
The discretization of $\kappa$ and $c^{(j)}$ with respect to $x_i$ (being equidistant by construction) then yields the discretized $\tilde{c}^{(j)},\tilde{\kappa}\in \R^N$ such that minimizing the discretized functional
\begin{equation}
    \tilde{J}_\text{pulsed}(\tilde{\kappa}) = \sum_j \| \tilde{c}^{(j)} \ast \tilde{\kappa} - \tilde{g}^{j} \|_2^2
\end{equation}
yields
 \begin{equation}\label{eq:kernel_extimate_pulsed}
     \hat{\tilde{\kappa}}_k = \frac{\sum_j \overline{\hat{\tilde{c}}^{(j)}_k} {\hat{\tilde{g}}^{(j)}_k}}{\sum_j \overline{\hat{\tilde{c}}^{(j)}_k} {\hat{\tilde{c}}^{(j)}_k}} 
 \end{equation}
 analogously to the sinusoidal case in \eqref{eq:kernel_extimate_sin}.

\begin{figure}
     \centering

     \begin{minipage}{0.45\textwidth}
          \includegraphics[width=\textwidth]{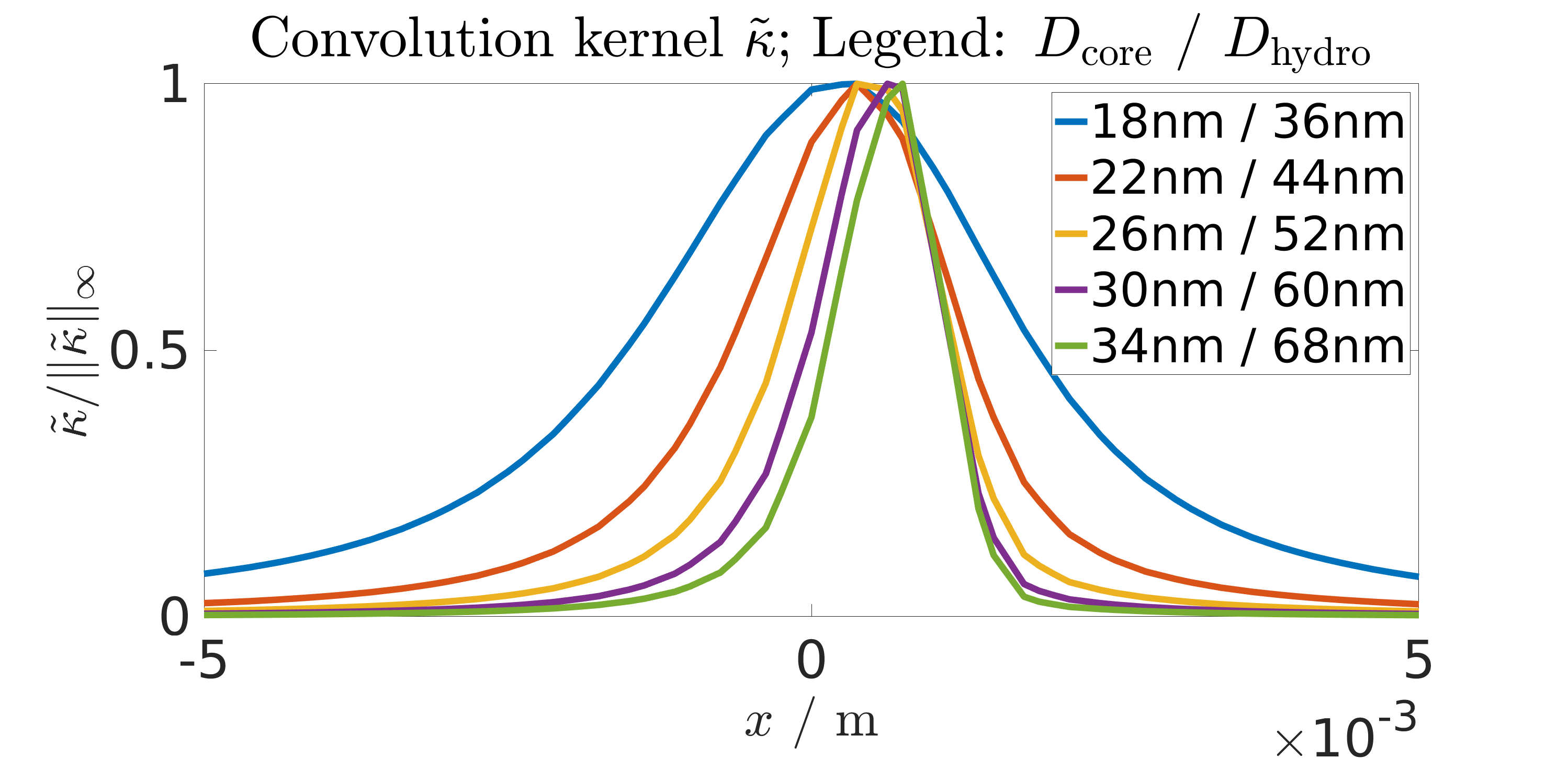}
     \end{minipage}
          \begin{minipage}{0.45\textwidth}
          \includegraphics[width=\textwidth]{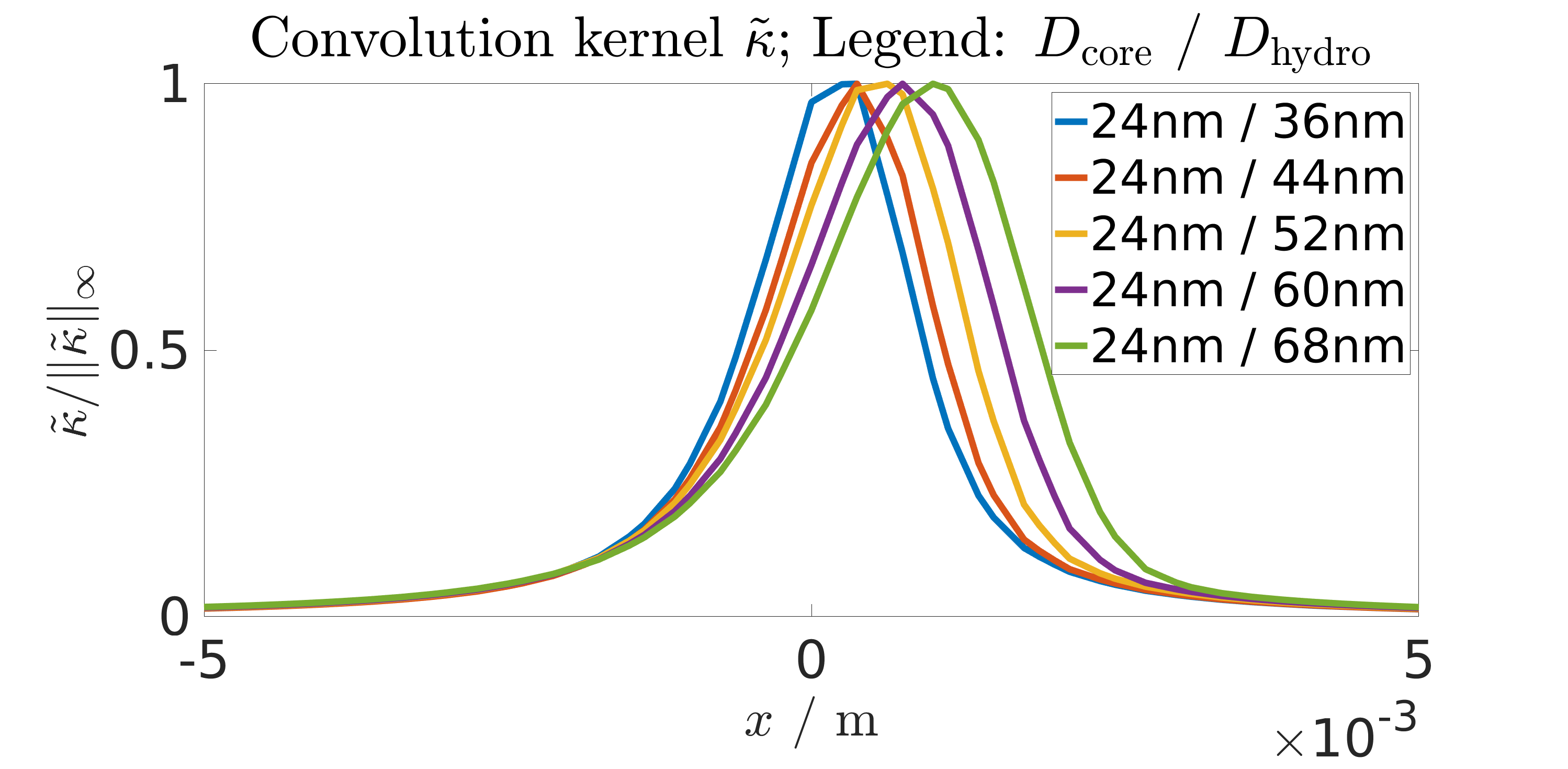}
     \end{minipage}
          \begin{minipage}{0.45\textwidth}
          \includegraphics[width=\textwidth]{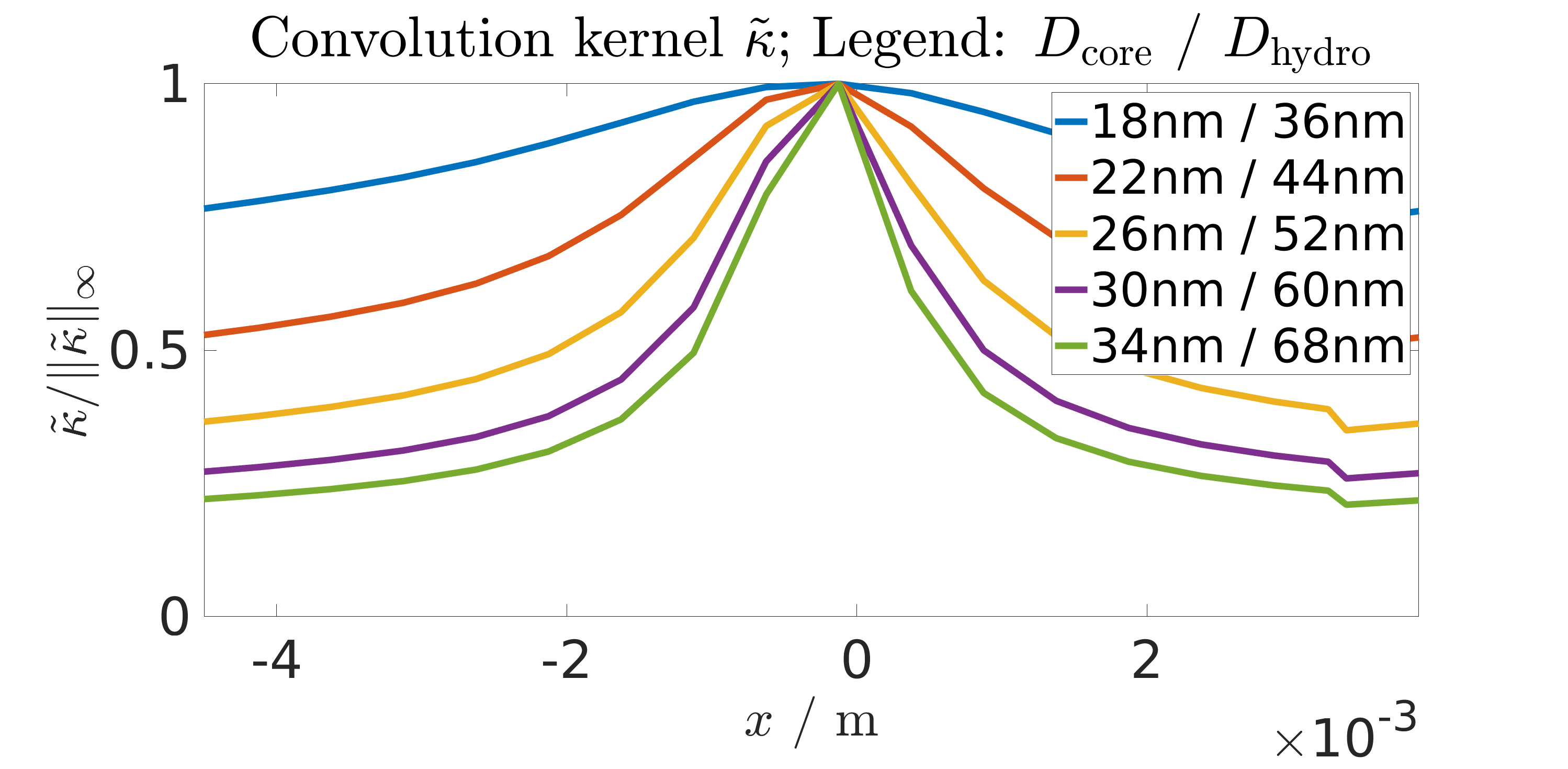}
     \end{minipage}
     \caption{
     Fitted convolution kernels $\tilde{\kappa}$ for various particle diameters (core and hydrodynamic) for sinusoidal (top) and pulsed (bottom) excitation.}
     \label{fig:x-space-kernel-fits}
 \end{figure}

 \paragraph{And finally, we discuss results obtained by the previous methods} 
 We exploit the computational framework from Section \ref{sec:computational_framework} to generate concentration-voltage tuples from the Brownian rotation model, which are then used to obtain the desired convolution kernels $\tilde{\kappa}$ as described in the previous part of this subsection. 
 For this we parameterize the excitation patterns according to the work in \cite{tay2019pulsed} to enable a qualitative comparison (particularly to \cite[Fig. 6]{tay2019pulsed}). 
 The used parameters are summarized in Table \ref{tab:pulsed_paramters}.
 One concentration-voltage tuple $(c^{(1)},v^{(1)})$ with  $c^{(1)}$ being a discrete delta sample placed at the origin is used to determine the kernel $\tilde{\kappa}$ according to \eqref{eq:kernel_extimate_sin} and \eqref{eq:kernel_extimate_pulsed}.
 The result for different tuples of core and hydrodynamic particle diameters are illustrated in Figure \ref{fig:x-space-kernel-fits}.
 
 From these findings, we can make three major observations confirming  findings from the literature:
 \begin{itemize}
     \item For the pulsed sequence (Figure \ref{fig:x-space-kernel-fits} (bottom)) increasing the particle diameter results in sharper convolution kernels being beneficial for image reconstructions. This finding is in line with the qualitative behavior reported in \cite{tay2019pulsed}. 
     \item For the sinusoidal case when simultaneously increasing hydrodynamic and core diameter (Figure \ref{fig:x-space-kernel-fits} (top, left)), the kernel becomes sharper again. This predicts the qualitative behavior in \cite[Fig. 6]{tay2019pulsed} up to the diameter 24.4~nm only.
     \item Keeping the core diameter fixed in the sinusoidal case (Figure \ref{fig:x-space-kernel-fits} (top, right)) and increasing the hydrodynamic diameter causes a less sharp convolution kernel being additionally shifted and sheared. This is in line with the qualitative behavior in \cite[Fig. 6]{tay2019pulsed} reported for diameters ranging from 24.4~nm to 32.1~nm.
 \end{itemize}

 \begin{table}
\centering
 \begin{tabular}{ll|l}
 \textbf{Parameter} & & \textbf{Value}\\ 

 \midrule
 Magnetic permeability & $\mu_0$ & $4\pi\times 10^{-7} \text{ H/m}$\\
 Boltzmann constant & $k_\mathrm{B}$ & $1.38064852\times 10^{-23} \text{ J/K}$\\

\midrule 
\multicolumn{3}{l}{{\it Scanner (sinusoidal)}} \\
\midrule
 Excitation frequency & $f$& $25000 \text{ Hz}$\\
 Excitation amplitudes & $A$& $0.02\text{ T} /\mu_0$\\
 Gradient strength & $G$ & $7\text{ T/m} /\mu_0$\\
 \midrule
 \multicolumn{3}{l}{{\it Scanner (pulsed)}} \\
\midrule
 Excitation frequency & $f_\mathrm{pulsed}$& $2500 \text{ Hz}$\\
 Excitation amplitudes & $A_\mathrm{pulsed}$& $0.001\text{ T} /\mu_0$\\
 Gradient strength & $G$ & $7\text{ T/m} /\mu_0$\\
 \midrule
\multicolumn{3}{l}{{\it Particle (spherical shape); Brownian rotation}} \\
 \midrule
  Temperature & $T_\textrm{B}$ & $293 \text{ K}$ \\
  Sat. magnetization & $M_{\mathrm{S}}$ & $474000 \text{ J/m$^3$/T}$\\
  Particle core diameter & $D_\mathrm{core}$ & variable \\
  Particle core volume & $V_\mathrm{C}$ & $1/6 \pi D_\mathrm{core}^3$ \\
  Particle core diameter & $D_\mathrm{hydro}$ & variable \\
  Particle hydrodynamic volume & $V_\mathrm{H}$ & $1/6 \pi D_\mathrm{hydro}^3$  \\
  Dynamic viscosity  & $\eta$ & $1.0 \times 10^{-5} \text{ Pa s} $\\
 
 \end{tabular}
 
 \caption{Physical parameters used in the simulations. }
\label{tab:pulsed_paramters}
 \end{table}

\section{Discussion}
\label{sec:discussion}

In this work, we presented two numerical algorithms that allow precise individual simulation of two mechanisms of nanoparticle dynamics in an applied magnetic field. We showed how computation can be sped up by introducing an approximation, numerically examined the accuracy of the proposed methods, and presented two promising applications to parameter identification problems.

Both methods have their individual benefits. For particle diameters up to about $\SI{30}{nm}$, the spherical harmonics method yields better performance. However, for large diameters and anisotropies, the method becomes unstable, ultimately failing to converge. Although we have not examined it in this work, we have observed anecdotally that time-varying easy axes only worsen this problem.

The finite volume method is slower for small particle diameters, but exhibits superior performance for large diameters and anisotropy constants. Furthermore, it is much more flexible with regards to modifications in the underlying equation; structurally different advection terms can be realized with no modification to the discretization procedure. This makes it possible, for example, to introduce more terms to the effective magnetic field in the N\'{e}el model or to consider cubic or hexagonal anisotropy, or a combination.

In the numerical evaluations in Section \ref{sec:comp_eval}, we only considered N\'{e}el relaxation since it poses more problems computationally. For evaluating Brownian rotation, computation times as well as stability w.r.t. parameter choice was significantly better behaved, and should lead to less computational problems in practice.

The two parameter identification problems have yielded promising results. Concerning the dictionary fit to the MPS data (E1), the simulated dynamics were unable to explain the measured behavior for a $\ang{90}$ angle between the assumed easy axis and the magnetic field. This may be due to the cubic magnetocrystalline anisotropy that iron oxide nanoparticles possess \cite{shasha2020}. It is unclear how the total anisotropy can be described and how the crystal lattice, the particle shape and the particle surface influence it in practice, see e.g. \cite{usov2012anisotropy}. If the assumption of uniaxial anisotropy is sufficient or the cubic structure must be taken into account remains to be answered in future work.

Regarding the second parameter identification problem we were able to obtain qualitative results by the simulation framework, which are in line with the experimental findings reported in the literature \cite{tay2019pulsed}. 
However, a sole Brownian rotation in the sinusoidal excitation case is not sufficient to explain the observed qualitative behavior in the experiments. Some adaptations in the balance between the diffusion and the advection term in the Fokker-Planck equation (e.g., fixing the core diameter and changing the hydrodynamic diameter) allow for the generation of the desired effect. From a modeling perspective one can expect that the N\'{e}el rotation model also requires some analogous adaptations to reproduce the sinusoidal case. In contrast, for the pulsed sequence the sole Brownian model is sufficient. 
The need for adaptations in the sinusoidal case could be related to an interplay of Brownian and N\'{e}el rotation due to their coupling by the particle anisotropy. Exploring this direction and extending the computational framework to the fully coupled Brown-N\'{e}el rotation case remains future work. 

Furthermore, while simulations for a single pixel in a field of view as in MPI can be obtained in a reasonable time, simulations of N\'{e}el rotation for a 3D field of view, in order to obtain a 3D system matrix, are still not feasible. In order to tackle this problem, further simplifications to the equation, or different solution methods, have to be developed in the future.

Nevertheless, we believe that this work constitutes a step further towards model-based reconstruction in MPI and towards a deeper understanding of nanoparticle dynamics in a qualitative as well as quantitative manner.

\bibliography{mypub,ref}

\appendix 
\section{Calculation of matrix elements for the SH method}\label{sec:appendix_SH}\label{sec:appendix}
In the following, we want to calculate the matrix elements $A(Y_j,Y_i)$ for the special advection term $\mathbf{b} = p_1 H \times m + p_2 (m \times H) \times m$ for (physical) parameters $p_1, p_2 > 0$. Note that the letter $m$ will be used to denote the magnetic moment direction as well as one of the integer spherical harmonic indices. Whenever $m$ is used together with a cross product, it denotes the magnetic moment direction.

Since calculations involving spherical harmonics are easiest in spherical coordinates, we first need to do a coordinate transformation for the integral of interest, i.e. $\int_S (\mathbf{b}\cdot \nabla Y_i)Y_j^*\, dx$.
Since 
\begin{align}
    \nabla Y_i &= \frac{\partial Y_i}{\partial \theta} \hat{\theta} + \frac{1}{\sin \theta} \frac{\partial Y_i}{\partial \varphi} \hat{\varphi}\\
    &= \frac{\partial Y_i}{\partial \theta} \begin{pmatrix}\cos \theta \cos \varphi\\
    \cos \theta \sin \varphi\\
    -\sin \theta 
    \end{pmatrix} + \frac{1}{\sin \theta} \frac{\partial Y_i}{\partial \varphi} \begin{pmatrix} -\sin \varphi\\
    \cos \varphi\\
    0\end{pmatrix},
\end{align}
where $\hat{\theta}$ and $\hat{\varphi}$ are the unit vectors in $\theta$ and $\varphi$ direction, respectively, and
\begin{align}
    H \times m &= \begin{pmatrix} H_2 m_3 - H_3 m_2\\
    H_3 m_1 - H_1 m_3 \\
    H_1 m_2 - H_2 m_1 \end{pmatrix}\\
    &= \begin{pmatrix} H_2 \cos \theta - H_3 \sin \theta \sin \varphi\\
    H_3 \sin \theta \cos \varphi - H_1 \cos \theta \\
    H_1 \sin \theta \sin \varphi - H_2 \sin \theta \cos \varphi\end{pmatrix},
\end{align}
we finally obtain, rewriting all terms involving $\varphi$ in terms of complex exponentials: 
\begin{equation}\label{hm}
\begin{aligned}
(H \times m) \cdot \nabla Y &= \frac{1}{2} (H_2 + iH_1) e^{i \varphi}\left( \frac{\partial}{\partial \theta} + i \cot \theta \frac{\partial }{\partial \varphi}\right) Y \\
    &-\frac{1}{2} (H_2-iH_1) e^{-i\varphi} \left( -\frac{\partial}{\partial \theta} + i \cot \theta \frac{\partial}{\partial \varphi} \right) Y\\
    &+ H_3 \frac{\partial Y}{\partial \varphi}.
    \end{aligned}
\end{equation}
Analogously, we get
\begin{align}
    (m\times H) \times m = \begin{pmatrix}1 - \sin ^2 \theta \cos ^2 \varphi & -\sin^2\theta \sin \varphi \cos \varphi & - \cos \theta \sin \theta \cos \varphi \\
    -\sin^2\theta \cos \varphi \sin \varphi & 1-\sin^2\theta \sin^2 \varphi & -\cos \theta \sin \theta \sin \varphi \\
    -\sin \theta \cos \theta \cos \varphi & -\sin \theta \cos \theta \sin \varphi & 1 - \cos^2\theta\end{pmatrix} \begin{pmatrix} H_1 \\ H_2 \\ H_3 \end{pmatrix}.
\end{align}
A rather lengthy calculation then yields:
\begin{equation}\label{mhm}
    \begin{aligned}
    ((m \times H) \times m )\cdot \nabla Y&= \frac{1}{2}(H_1-iH_2)e^{i\varphi} \left( \cos \theta \frac{\partial}{\partial \theta} + i \frac{1}{\sin \theta} \frac{\partial}{\partial \varphi} \right) Y \\
    &- \frac{1}{2}(H_1 + iH_2) e^{-i\varphi}\left(-\cos \theta \frac{\partial}{\partial \theta} + i \frac{1}{\sin \theta} \frac{\partial}{\partial \varphi}\right) Y \\
    &- H_3 \sin \theta \frac{\partial Y}{\partial \theta}.
    \end{aligned}
\end{equation}
Our goal is now to rewrite \eqref{hm} and \eqref{mhm} in such a way, that we can immediately exploit the orthogonality of the spherical harmonics without having to numerically calculate any integral. To this end, we need the following properties of the non-normalized spherical harmonics $Y^m_l$ [Weizenecker2018, Dunster2010]:
\begin{align}
    \cos (\theta) Y^m_l &= \frac{1}{2l+1}\left((l+m)Y^m_{l-1} + (l+1-m)Y^m_{l+1}\right)\label{cos}\\
    \sin (\theta) Y^m_l &= \frac{1}{2l+1}(Y^{m+1}_{l-1} - Y^{m+1}_{l+1})e^{-i\varphi}\\
    \sin (\theta) Y^m_l &= \frac{1}{2l+1}\left((l-m+1)(l-m+1)Y^{m-1}_{l+1} - (l+m-1)(l+m)Y^{m-1}_{l-1}\right) e^{i\varphi}\label{sin}\\
    \sin (\theta) \frac{\partial Y^m_l}{\partial \theta} &= \frac{1}{2l+1}\left(l(l-m+1)Y^m_{l+1} - (l+1)(l+m)Y^m_{l-1}\right)\\
    \frac{\partial Y^m_l}{\partial \varphi} &= imY^m_l\label{dphi}\\
    (Y^m_l)^* &= Y^{-m}_l\\
    \int_S Y^m_l (Y^M_L)^*\, dx &= \int_S Y^m_l Y^{-M}_l\, dx = \frac{4\pi (-1)^M}{2L+1} \delta_{L,l}\delta_{M,m}.\label{normalization}
\end{align}
Using the identities \eqref{cos}-\eqref{dphi}, we can calculate the following:
\begin{align}
    e^{i\varphi}\left( \frac{\partial }{\partial \theta} + i\cot \theta \frac{\partial }{\partial \varphi} \right) Y^m_l &= Y^{m+1}_l\\
    e^{-i\varphi}\left(-\frac{\partial}{\partial \theta} + i\cot \theta \frac{\partial}{\partial \varphi}\right) Y^m_l &= (l+m)(l-m+1)Y^{m-1}_l\\
    e^{i\varphi} \left( \cos \theta \frac{\partial}{\partial \theta} + i \frac{1}{\sin \theta} \frac{\partial}{\partial \varphi}\right) Y^m_l &= \frac{1}{2l+1}\left((l+1)Y^{m+1}_{l-1} + lY^{m+1}_{l+1}\right)\\
    e^{-i\varphi}\left( -\cos \theta \frac{\partial}{\partial \theta} + i \frac{1}{\sin\theta} \frac{\partial}{\partial \varphi} \right) Y^m_l &= \left(\frac{l(l-m+1)(l+m-1)}{2l+1} +m(l+m-1)\right) Y^{m-1}_{l-1} \\
    &+ \frac{l(l-m+1)(l-m+2)}{2l+1} Y^{m-1}_{l+1}\nonumber
\end{align}
With these results, we can rewrite \eqref{hm} and \eqref{mhm} entirely in terms of a combination of a finite number of spherical harmonics with coefficients that are independent of $m$:
\begin{equation}\label{expansionmhm}
    \begin{aligned}
    \left((m\times H)\times m\right)\cdot \nabla Y^m_l &= \frac{1}{2}(H_1-iH_2) \frac{1}{2l+1}\left((l+1)Y^{m+1}_{l-1} + lY^{m+1}_{l+1}\right)\\
    &-\frac{1}{2}\frac{1}{2l+1} (H_1+iH_2) l(l-m+1)\left((l+m-1)Y^{m-1}_{l-1} + (l-m+2)Y^{m-1}_{l+1}\right) \\
    &-\frac{1}{2}(H_1+iH_2)m(l+m-1)Y^{m-1}_{l-1} \\
    &- H_3 \frac{1}{2l+1}\left(l(l-m+1)Y^m_{l+1} - (l+1)(l+m)Y^m_{l-1}\right),
    \end{aligned}
\end{equation}
\begin{equation}\label{expansionhm}
    \begin{aligned}
    (H \times m) \cdot \nabla Y^m_l &= \frac{1}{2}(H_2+iH_1)Y^{m+1}_l \\
    &- \frac{1}{2}(H_2-iH_1)(l+m)(l-m+1)Y^{m-1}_l + H_3 imY^m_l
    \end{aligned}
\end{equation}
Abstractly, we have found coefficients $a_{ij}(m,l)$ such that
\begin{align}\label{gradient-rep}
    (\mathbf{b}\cdot \nabla Y^m_l) = \sum_{i,j=-2}^2 a_{ij}(m,l)Y^{m+i}_{l+j}.
\end{align}
Now, let us return to the Fokker-Planck equation. Taking the product of the equation with a specific spherical harmonic function $Y^{-q}_r$ and integrating over $S$, we obtain
\begin{align*}
    \int \frac{\partial f}{\partial t}Y^{-q}_r \, dx &= \int \Delta f\, Y^{-q}_r\, dx -\int \diva \left( \mathbf{b}\cdot f\right) Y^{-q}_r \, dx.
\end{align*}
Next, we insert the expansion $f(t) = \sum_{l,m}C^m_l(t)Y^m_l$:
\begin{align*}
    \sum_{l,m}\frac{\partial}{\partial t}C^m_l(t) \int Y^m_l Y^{-q}_r\, dx &= \sum_{l,m} C^m_l(t)\int \Delta Y^m_l \, Y^{-q}_r\, dx- \int \diva \left(\mathbf{b}\cdot \nabla  Y^m_l C^m_l(t) \right) Y^{-q}_r \, dx\\
    &=-r(r+1)C^q_r(t)\int Y^q_r Y^{-q}_r \, dx + \sum_{l,m}C^m_l(t) \int \left( \mathbf{b}\cdot \nabla Y^{-q}_r\right) Y^m_l \, dx.
\end{align*}
Inserting \eqref{gradient-rep}, taking into account \eqref{normalization}, we arrive at
\begin{align*}
    \frac{\partial}{\partial t}C^q_r(t) &= -r(r+1)C^q_r(t) + \sum_{i,j=-2}^2 a_{ij}(-q,r) \frac{(-1)^i (2r+1)}{2(r+j)+1} C^{q-i}_{r+j} (t).
\end{align*}
We point out that due to the spherical harmonics being non-normalized as well as them being complex, a sign flip $q \to -q$ in the calculated coefficients $a_{ij}$ as well as the normalization factors $\frac{(-1)^i (2r+1)}{2(r+j)+1}$ have to be taken into account.\\
This allows us to write down the discretized system of ODEs for each coefficient $C^q_r(t),\, q=-r,\dots,r;\, r=0,\dots,N$ for arbitrary $N\in \mathbb{N}$ for the special advection term of the form $\mathbf{b} = p_1 H \times m + p_2 (m\times H)\times m$:
\begin{equation}\label{eq:fp_disc}
    \begin{aligned}
    \frac{\partial C^q_r}{\partial t} = &- r(r+1)C^q_r - i \frac{p_1}{2}\left( (H_1+iH_2) (r-q)(r+q+1) C^{q+1}_r + (H_1-iH_2) C^{q-1}_r + 2qH_3 C^q_r\right)\\
    &+ p_2 H_3 \left( \frac{(r+1)(r-q)}{2r-1}C^q_{r-1} - \frac{r(r+q+1)}{2r+3} C^q_{r+1}\right)\\
    &+p_2 (H_1+iH_2)\left( \frac{(r+1)(r-q)(r-q-1)}{4r-2}C^{q+1}_{r-1} + \frac{r(r+q+1)(r+q+2)}{4r+6} C^{q+1}_{r+1}\right)\\
    &+p_2 (H_1-iH_2) \left( -\frac{(r+1)}{4r-2} C^{q-1}_{r-1} - \frac{r}{4r+6} C^{q-1}_{r+1}\right).
    \end{aligned}
\end{equation}
This leaves the anisotropy part of the advection term, $\mathbf{b}_2 = p_3 \, (n \cdot m) n\times m + p_4 \, (n \cdot m) (m \times n) \times m$. This can be rewritten as
\begin{align*}
    \mathbf{b}_2 &= p_3 \left(n_1 \cos(\varphi)\sin(\theta)+n_2 \sin(\varphi)\sin(\theta) + n_3 \cos(\theta)\right) n \times m\\
    &+p_4\left(n_1 \cos(\varphi)\sin(\theta)+n_2 \sin(\varphi)\sin(\theta) + n_3 \cos(\theta)\right) (m \times n)\times m.
\end{align*}
Using \eqref{cos}-\eqref{sin}, we can immediately calculate the effect that the spherical unit vector components has on a non-normalized spherical harmonic function $Y^m_l$:
\begin{align*}
    \cos(\varphi)\sin(\theta)Y^m_l &= \frac{1}{2(2l+1)}\left(Y^{m+1}_{l-1} - Y^{m+1}_{l+1} + (l-m+1)(l-m+2)Y^{m-1}_{l+1} - (l+m-1)(l+m)Y^{m-1}_{l-1}\right),\\
    \sin(\varphi)\sin(\theta)Y^m_l &= \frac{i}{2(2l+1)}\left(Y^{m+1}_{l+1} - Y^{m+1}_{l-1} + (l-m+1)(l-m+2)Y^{m-1}_{l+1} - (l+m-1)(l+m)Y^{m-1}_{l-1}\right),\\
    \cos(\theta)Y^m_l &= \frac{1}{2l+1} \left( (l+m)Y^m_{l-1} + (l+1-m)Y^m_{l+1}\right).
\end{align*}
With this, we can calculate the coefficients $a_{ij}(q,r)$ as in \eqref{gradient-rep} by taking the results \eqref{expansionmhm}, \eqref{expansionhm}, replacing $H$ by $n$ and applying the operators as above. This finally leads to an ODE system of the form
\begin{align}
        \frac{\partial C^q_r}{\partial t} = \sum_{q'=q-2}^{q+2}\sum_{r'=r-2}^{r+2}\gamma^{q,q'}_{r,r'}(t) C^{q'}_{r'}(t), \qquad q=-r,\dots,r,\quad r=0,\dots,\infty,,
\end{align}
where the coefficients $\gamma^{q'}_{r'}$ corresponding to $C^{q'}_{r'}$ are summarized in table \ref{tablecoeffs} (for the anisotropy part) and equation \eqref{eq:fp_disc} (for the diffusion and non-anisotropy part).
\newpage
\begin{table}
\begin{center}\setlength\extrarowheight{5pt}\large{
\caption{Coefficients corresponding to the anisotropy term}\label{tablecoeffs}
    \begin{tabular}{||l | l||}
        $C^{m'}_{l'}$ & $\gamma^{m,m'}_{l,l'}$\\
        \hline
        $C^{m-2}_{l-2}$ & $\frac{p_4(l+1)}{4(2l-3)(2l-1)}(n_1-in_2)^2$\\
        $C^{m-1}_{l-2}$ & $\frac{-p_4(l+1)(l-m)}{(2l-3)(2l-1)}n_3(n_1-in_2)$\\
        $C^m_{l-2}$ & $\frac{-p_4(l+1)(l-m)(l-m-1)}{2(2l-3)(2l-1)}(n_1^2+n_2^2-2n_3^2)$\\
        $C^{m+1}_{l-2}$ & $\frac{p_4(l+1)(l-m)(l-m-2)(l-m-1)}{(2l-3)(2l-1)}n_3(n_1+in_2)$\\
        $C^{m+2}_{l-2}$ & $\frac{p_4(l+1)(l-m)(l-m-3)(l-m-2)(l-m-1)}{4(2l-3)(2l-1)}(n_1+in_2)^2$\\
        $C^{m-2}_{l-1}$ & $\frac{ip_3}{4(2l-1)}(n_1-in_2)^2$\\
        $C^{m-1}_{l-1}$ & $\frac{ip_3(2m-l-1)}{2(2l-1)}n_3(n_1-in_2)$\\
        $C^m_{l-1}$ & $\frac{ip_3 m (l-m)}{2(2l-1)}(n_1^2+n_2^2-2n_3^2)$\\
        $C^{m+1}_{l-1}$ & $\frac{-ip_3(l-m)(l-m-1)(2m+l+1)}{2(2l-1)}n_3(n_1+in_2)$\\
        $C^{m+2}_{l-1}$ & $\frac{-ip_3(l-m)(l-m-1)(l-m-2)(l+m+1)}{4(2l-1)}(n_1+in_2)^2$\\
        $C^{m-2}_l$ & $\frac{-p_4}{4(2l-1)(2l+3)}3(n_1-in_2)^2$\\
        $C^{m-1}_l$ & $\frac{-p_4}{2(2l-1)(2l+3)}3(2m-1)n_3(n_1-in_2)$\\
        $C^m_l$ & $\frac{-p_4(l^2+l-3m^2)}{2(2l-1)(2l+3)}(n_1^2+n_2^2-2n_3^2)$\\
        $C^{m+1}_l$ & $\frac{-p_4(2m+1)(l-m)(l+m+1)}{2(2l-1)(2l+3)}3n_3(n_1+in_2)$\\
        $C^{m+2}_l$ & $\frac{-p_4(l-m)(l-m-1)(l+m+1)(l+m+2)}{4(2l-1)(2l+3)}3(n_1+in_2)^2$\\
        $C^{m-2}_{l+1}$ & $\frac{-ip_3}{4(2l+3)}(n_1-in_2)^2$\\
        $C^{m-1}_{l+1}$ & $\frac{-ip_3(2m+l)}{2(2l+3)}n_3(n_1-in_2)$\\
        $C^m_{l+1}$ & $\frac{ip_3m(l+m+1)}{2(2l+3)}(n_1^2+n_2^2-2n_3^2)$\\
        $C^{m+1}_{l+1}$ & $\frac{ip_3(2m-l)(l+m+1)(l+m+2)}{2(2l+3)}n_3(n_1+in_2)$\\
        $C^{m+2}_{l+1}$ & $\frac{ip_3(l-m)(l+m+1)(l+m+2)(l+m+3)}{4(2l+3)}(n_1+in_2)^2$\\
        $C^{m-2}_{l+2}$ & $\frac{-p_4 l}{4(2l+3)(2l+5)}(n_1-in_2)^2$\\
        $C^{m-1}_{l+2}$ & $\frac{-p_4 l(l+m+1)}{(2l+3)(2l+5)}n_3(n_1-in_2)$\\
        $C^m_{l+2}$ & $\frac{p_4 l (l+m+1)(l+m+2)}{2(2l+3)(2l+5)}(n_1^2+n_2^2-2n_3^2)$\\
        $C^{m+1}_{l+2}$ & $\frac{p_4l(l+m+1)(l+m+2)(l+m+3)}{(2l+3)(2l+5)}n_3(n_1+in_2)$\\
        $C^{m+2}_{l+2}$ & $\frac{-p_4l(l+m+1)(l+m+2)(l+m+3)(l+m+4)}{4(2l+3)(2l+5)}(n_1+in_2)^2$
    \end{tabular}}
\end{center}
\end{table}

\end{document}